\documentclass[sigconf, pageno, screen, authorversion]{acmart}

\pdfoutput=1
\usepackage{subfig}
\usepackage{lipsum}
\usepackage{tabularx}
\usepackage{graphicx}
\usepackage{xcolor}
\usepackage{subcaption}
\usepackage{enumitem}
\usepackage{lineno}
\usepackage{tablefootnote}
\usepackage{xspace}

\renewcommand\footnotetextcopyrightpermission[1]{}

\AtBeginDocument{%
  }

\colorlet{blue}{black}

\copyrightyear{2025}
\acmYear{2025}
\setcopyright{rightsretained}
\acmConference[e-Energy '25]{The 16th ACM International Conference on Future and Sustainable Energy Systems}{June 17--20, 2025}{Rotterdam, Netherlands} 
\acmBooktitle{The 16th ACM International Conference on Future and Sustainable Energy Systems (e-Energy '25), June 17--20, 2025, Rotterdam, Netherlands}
\acmISBN{978-1-4503-XXXX-X/18/06}

\newcommand{\adaml}[1]{  
	{\textcolor{purple}{(Adam L. says:  #1)}}{}}

\newcommand{\sref}[2]{\hyperref[#2]{#1 \ref{#2}}}
	
\definecolor{OliveGreen}{rgb}{0,0.6,0}

\newcommand{\toolkit}{\texttt{SunSight}\xspace} 
\begin{document}

% \title{The Surprising Carbon Inefficiency Caused by Inequitable Distribution of Rooftop Solar Installations in the US}
% \title[Missing the Solar Mark]{Missing the Solar Mark: The Hidden Carbon Cost of Inequitable Residential Solar Installations}
\title[Lost in Siting]{Lost in Siting: The Hidden Carbon Cost \\of Inequitable Residential Solar Installations}

\author{Cooper Sigrist}
\email{csigrist@umass.edu}
\affiliation{%
  \institution{University of Massachusetts Amherst}
  \city{Amherst}
  \state{Massachusetts}
  \country{USA}
}

\author{Adam Lechowicz}
\email{alechowicz@umass.edu}
\affiliation{%
  \institution{University of Massachusetts Amherst}
  \city{Amherst}
  \state{Massachusetts}
  \country{USA}
}

\author{Jovan Champ}
\email{champ@hartford.edu}
\affiliation{%
  \institution{University of Hartford}
  \city{Hartford}
  \state{Connecticut}
  \country{USA}}

\author{Noman Bashir}
\email{nbashir@mit.edu}
\affiliation{%
  \institution{Massachusetts Institute of Technology}
  \city{Cambridge}
  \state{Massachusetts}
  \country{USA}
}

\author{Mohammad Hajiesmaili}
\email{hajiesmaili@cs.umass.edu}
\affiliation{%
  \institution{University of Massachusetts Amherst}
  \city{Amherst}
  \state{Massachusetts}
  \country{USA}
}

\renewcommand{\shortauthors}{Sigrist et al.}

\begin{abstract}
The declining cost of solar photovoltaics (PV) combined with strong federal and state-level incentives have resulted in a high number of residential solar PV installations in the US.
However, these installations are concentrated in particular regions, such as California, and demographics, such as high-income Asian neighborhoods. 
This inequitable distribution creates an illusion that further increasing residential solar installations will become increasingly challenging if it is not already prohibitive. 
Furthermore, while the inequity in solar installations has received attention, no prior comprehensive work has been done on understanding whether our current trajectory of residential solar adoption is energy- and carbon-efficient.

In this paper, we reveal the hidden energy and carbon cost of the inequitable distribution of existing installations. 
Using US-based data on carbon offset potential—the amount of avoided carbon emissions from using rooftop PV instead of electric grid energy—and the number of existing solar installations, we surprisingly observe that locations and demographics with a higher carbon offset potential have fewer existing installations.
For instance, neighborhoods with relatively higher black population have 7.4\% higher carbon offset potential than average but 36.7\% fewer installations; lower-income neighborhoods have 14.7\% higher potential and 47\% fewer installations; Republican-leaning states have 23.8\% higher potential and 60.8\% fewer installations. 
We propose several equity- and carbon-aware solar siting strategies that prioritize developing solar in certain areas based on their characteristics -- these strategies may inform, for example, the development of targeted incentives.  
In evaluating these strategies, we develop \toolkit, a toolkit that combines simulation/visualization tools and our relevant datasets, which we are releasing publicly.
Our projections show that an multi-objective siting strategy can address two problems at once -- namely, it can improve societal outcomes in terms of distributional equity and simultaneously improve the carbon-efficiency (i.e., climate impact) of current installation trends by up to 39.8\%.

\end{abstract}

\maketitle

\section{Introduction}
\label{sec:intro}
In 2021, the United States announced a goal to reduce national carbon emissions by 50-52\% from its 2005 levels by 2030. Further, it announced plans to reach 100\% carbon-free energy by 2035~\cite{The_White_House_2021}. These ambitious decarbonization goals come with substantial investments in new technologies and infrastructure, causing a new green industry boom. 
In 2022, global solar photovoltaics (PV) accounted for three-quarters of all renewable capacity additions worldwide, with the International Energy Agency (IEA) predicting a doubling of global solar PV capacity in the next four years and a tripling in the next six years to achieve net zero emissions~\cite{Iea}. Investment in PV, and the total number of new installations, is at an all-time, yearly high~\cite{SPE:23, DOESolarCost, Lemay:23}. This is likely because photovoltaics have two key advantages over other carbon-free energy options:

\begin{figure}[t]
\vspace{1.5em}
    \centering
    \includegraphics[width=\columnwidth]{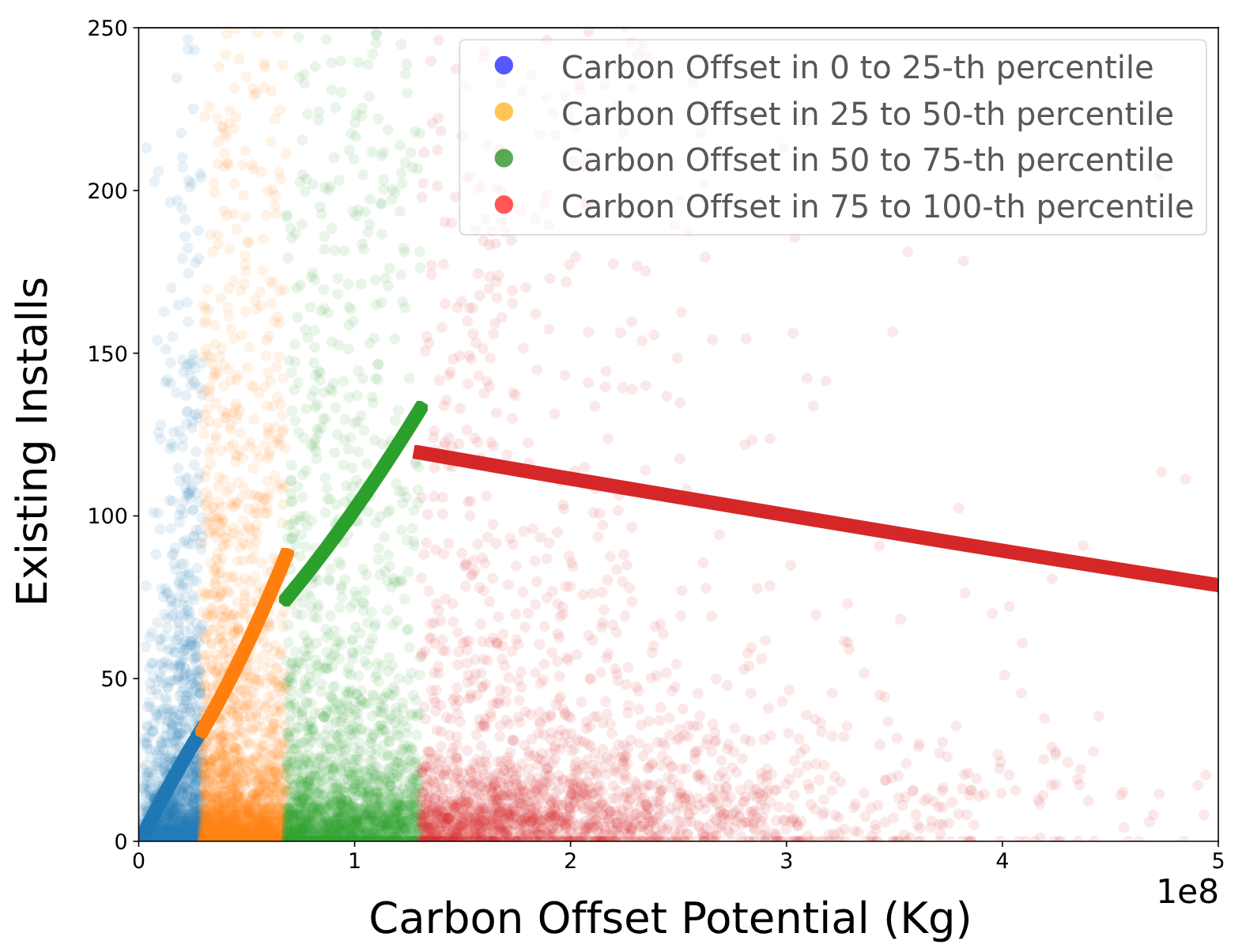}
    \vspace{-1em}
    \caption{{\color{blue}The estimated potential carbon offset (defined in \autoref{sec:prob}) of building all viable rooftop solar panels plotted against the number of current rooftop installs for each ZIP code covered by Google's Project Sunroof. ZIP codes with > 250 existing installs or potential carbon offset > 800,000 metric tons are not shown for legibility, but are used to fit the trend lines. The ZIP codes are split along the quartiles of their carbon offset and each group is fit by a quadratic function to minimize LSE. Pearson correlation coefficients (PCC) for the first (blue, 0-25 percentile) to fourth quartile are 0.23, 0.12, 0.07, and -0.03 respectively.}}
     \vspace{-0.5cm}
    \label{fig:co_vs_existing_intro}
\end{figure}

\noindent \textbf{Cost Efficiency. \ } Levelized cost of energy (LCOE) is a standard metric used to compare electricity-generating technologies that consider the lifetime costs of a particular generator divided by the amount of energy produced over its expected lifespan.  As of 2024, utility-scale photovoltaics have an LCOE of between \$39-62 USD per MWh (megawatt-hour), while technologies such as offshore wind, geothermal, and hydropower have an LCOE anywhere between \$71-425 USD per MWh~\cite{ATB:2024}. The closest cost-efficient carbon-free generating source is land-based wind, with an LCOE between \$29-74 USD per MWh.

% The U.S. Energy Information Administration (EIA) reports that the cost of PV (solar panel with sun tracking in particular) installations per Kilowat (kW) of electricity was approximately \$1,327 USD \rev{or \$1,748 USD when accounting for energy storage} \mo{I cannot follow.}. This is comparable to the next most cost-efficient of the green energy choices, wind turbines, which have a cost efficiency of \$1,718/kW.

\noindent \textbf{Installation Flexibility. \ } While the last decade has seen significantly reducing costs for PV panels due to economies of scale, solar PV also comes with an unprecedented amount of spatial flexibility in their installations.  In contrast to space-constrained technologies such as wind and geothermal, rooftop PV installations have emerged around the globe, bringing a new paradigm of \textit{distributed energy resources} (DERs) to the forefront.
While residential rooftop PV is more costly than utility scale installations (LCOE between \$134-208 USD per MWh~\cite{ATB:2024}), it is an attractive option for homeowners and building owners who seek to reduce their energy costs or even sell energy back to the utility under feed-in tariffs~\cite{Peerapong2014}.
Furthermore, to incentivize adoption of rooftop PV, governments around the world have offered incentives such as tax credits, subsidies, and feed-in tariffs~\cite{Matisoff2017}.

% -- in the U.S., individual states

% such as New York offer an additional 15\% tax credit (up to \$1000) and sales tax exemption,  though these vary in number of offerings and value}. 

% , the federal government typically offers incentives, such as in the form of Tax Credits \rev{(TC)}, which allow for, say,  

% While comparable, clean options such as wind turbines or geothermal generators require very specific geographical conditions to justify building, PV installations require much less space and infrastructure, as well as being practical in a much larger portion of the nation. 

% These advantages are most apparent in one setting in particular: residential rooftop installations. Photovoltaics, unlike other clean energy generation options, may be built on top of existing structures, particularly in residential districts, where no comparable clean energy generation could legally be built. Its cost advantage is also emphasized in this setting. Building on existing foundations reduces the cost, and because the owner of the foundational building can reduce their energy costs (or even make profit) they will fund, in part or whole, the construction of the panels. 
While residential PV has seen significant growth in the United States over the last 10 years, new residential installations were at a quarterly \textit{low} since 2022 as of Spring 2024, driven by rising interest rates~\cite{SEIA_2024}.  Furthermore, prior work~\cite{Sunter:19, OShaughnessy:20} has shown that the \textit{distribution} of installations throughout the United States can be considered \textit{inequitable} along lines of race and ethnicity and/or income.  This inequitable distribution, combined with the above, may suggest that further increases in residential PV capacity will become increasingly difficult to achieve if e.g., wealthy areas have become saturated and existing cost/incentive structures are not effective for other areas.

% \mo{a connection centered around the sentence ``This inequitable distribution creates an illusion that further increasing residential solar installations will become increasingly challenging if it is not already prohibitive. '' in abstract could be added here.}

% With such an uncontested use case, cost advantage, and incentive programs, a rapid adoption of rooftop PV installations is expected. New residential, rooftop installations in Spring 2024, however, were at a quarterly  since 2022, likely due to rising interest rates.   
% \rev{In this work, we will show that this is not, in fact, caused by a lack of opportunities to decarbonize by building new rooftop Photovoltaics but rather, in part, by the inequitable distribution of rooftop installations. }\mo{sounds like a bit of deviation from what you want to eventually build your story on. you may say that prior work shown this thing (inequitable distribution), and you want to reveal another consequence of that, which is carbon inefficiency. } \adaml{agreed}

The above possibility holds implications for climate goals as well.  Since the ultimate goal of e.g., government incentives is to reduce the carbon intensity (i.e., CO$_2$ emissions per unit of electricity) throughout society, if these incentives fail to increase adoption, climate goals may not be feasible.  Furthermore, since existing works show that environmental impacts associated with the electricity sector (e.g., power plant pollution) are disproportionately felt by marginalized communities~\cite{Cranmer2023}, we consider the \textit{intersection} between climate and equity goals.  

In particular, we investigate and reveal the \textit{carbon inefficiency} that results from an inequitable distribution of rooftop PV installations in the United States.
% additional consequences that result from this inequitable distribution, namely 
Although prior works have extensively considered equity in installation count and sunlight generation potential, there is limited understanding of the \textit{carbon efficiency} of existing solar installations throughout the United States.  To gain an understanding of this, however, one must consider additional data points, such as information about projected energy generation and existing electric grid infrastructure at a fine-grained granularity.  

\subsection{Contributions}
In this paper, we make the following contributions. 

First, we use Google's Project Sunroof API to collect data on current installations, sunlight generation potential, and potential carbon offset estimates for the entire United States alongside corresponding demographic data.  In addition to the demographics considered in prior works (race/ethnicity, income), we additionally consider data points of geography and political voting records to paint a more comprehensive picture of disparities in residential solar installations.

Second, using this data, we demonstrate \textit{carbon inefficiency} -- a fundamental \textit{misalignment} between the current trends of rooftop installations and the broader societal goal of \textit{decarbonization}.  In contrast to prior works, analyzing the carbon efficiency of existing and future potential rooftop PV installations requires a substantial amount of information, including estimates of the yearly energy generation for each installation and grid carbon intensity (i.e., CO$_2$ emissions per unit of energy) at a fine-grained granularity.  By obtaining suitable data for these and other factors from disparate sources, we show that demographic inequities observed in this, and prior, works are likely major detractors from carbon efficiency, which quantifies the amount of CO$_2$ emissions prevented by each individual installation.
As an example, in \autoref{fig:co_vs_existing_intro}, we show that for ZIP codes that are in the upper quartile (top 25\%) of potential carbon offset, higher potential offsets seem to paradoxically correspond to fewer installations. We hypothesize that this is evidence both of the existence of a large number of homes that are viable for \textit{carbon-efficient} rooftop PV, as well as an important gap in existing incentivization strategies. 

%extend observations of demographic inequity, finding additional disparities based on geography and political affiliation. 
% Using data from Googles
% Using potential carbon offset estimations from Google's Project Sunroof, we demonstrate another troubling trend. In

Third, beyond analyzing the existing installations, in \autoref{sec:strategies} we leverage data on potential future installations to design and simulate new \textit{siting strategies} that identify locations where rooftop PV has the greatest impact.  In \autoref{sec:evaluation}, we demonstrate a large gap between current (``status quo'') installation trends and alternative approaches that optimize the placement of future installations for carbon efficiency or demographic equity.  

Such strategies must be considered in the context of broader decarbonization goals -- for instance, we show that a strategy that optimizes for carbon offsets can achieve $71.3\%$ more carbon reductions than current installation trends with the same amount of rooftop PV.
We propose a multi-objective strategy that balances between multiple attributes (energy, carbon, equity) in siting to achieve the ``best of both worlds''.  This relatively simple strategy is able to achieve $94.6\%$ of the energy production benefits of current installation trends, while improving carbon reductions (and thus climate impact) by at least $39.8\%$.

% Compared to the Status Quo, even relatively simple policies (such as Round Robin over multiple objectives) are not only able to achieve a similar (94.6\%) energy capacity gain, but will improve carbon offsetting tremendously (39.8\%), remain geographically equitable, and substantially reduce racial- and economic-inequity.

% Our results show that while optimizing for carbon can improve the emissions outcomes of current trends by , our proposed equity-driven approaches can achieve at least $77.0\%$ as much carbon reduction as the carbon-optimized plan, and a $39.8\%$ improvement compared to current trends.

% We explore possible different causes of this issue and ultimately find that the inequity of rooftop solar installations is deeply connected to carbon-inefficient rooftop PV installation trends. We propose possible solutions to this problem such as a change to the federal incentivization policy and addition of demographic-specific incentives.

Last, to enable other researchers to build on our work and explore other deployment strategies, we will publicly release our data and software as the \toolkit toolkit.\footnote{\toolkit is available at \url{https://github.com/coopersigrist/SunSight}.} \toolkit includes our compiled and cleaned dataset, visualization tools, and an environment to evaluate different incentivization strategies for residential solar installation.

\section{Background and Motivation}
\label{sec:background}
% \adaml{placeholder from Noman's outline -- writing today}
This section presents background information about rooftop photovoltaics and intersecting dynamics that motivate our study.

\noindent\textbf{Residential Solar PV. \ }
Over the last 10-15 years, the residential solar industry has seen significant growth in the United States, primarily driven by rooftop installations.  Such installations leverage the existing structure of a building's roof (e.g., of a single-family home) to mount several dozen photovoltaic cells.  Each installation has an estimated lifetime of $25$ years.  As of the time of writing, photovoltaics on residential rooftops make up 97\% of the installed solar capacity in the United States~\cite{SEIA_2024}.  These systems are generally interconnected with the local distribution grid, allowing homeowners to ``sell energy back to the grid'' via, e.g., net-metering programs~\cite{NCCETC:2023:FiftyStates}.

\noindent\textbf{Solar Energy and Carbon Offset Potential. \ }
Energy generation potential for photovoltaics depends on several factors. 
First, the sun's irradiance is unique to each location, affecting the amount of sunlight reaching the ground, even under clear skies. 
Weather conditions, especially cloud cover, are unpredictable and vary widely even in a local area, further impacting power output. Additionally, local physical characteristics such as module orientation, tilt, and shading from objects like trees or buildings can also influence the amount of energy generated. 
These create significant differences in solar potential between even closely situated sites. %, even between closely situated sites, let alone sites distributed across the US.

{\color{blue}The carbon offset potential depends on \textit{both} the solar energy generation and the carbon intensity of the grid electricity that the solar energy will replace, as well as their respective timings. 
The carbon intensity of grid electricity varies by location due to the unique mix of energy sources used to meet regional demand, which fluctuates based on demand and weather conditions. Renewable sources (e.g., solar, wind, hydro) have low or zero carbon emissions but are non-dispatchable, meaning their output is uncontrollable and depends on external conditions. In contrast, fossil fuel-based generators (e.g., coal and natural gas) have higher carbon intensities and are used to stabilize the grid when renewable output fluctuates. The mix of these sources, influenced by a region’s generation capacity, resources, and climate, leads to different annual average carbon intensities across locations.} %\mo{too much details here, could cut if need space eventually.}

% Each grid region, managed by an Independent System Operator or Regional Transmission Organization, must balance demand using generators with different characteristics, including carbon emission rates. 

% \adaml{cutting for length}
% Regions with higher renewable penetration, such as California, experience more \textit{variability} in carbon intensity due to the fluctuations of renewables. 
% However, our work uses annual carbon offset potential values that vary by location and do not consider the temporal changes in carbon intensity and, therefore, carbon offset potential.

\noindent\textbf{Incentives. \ }
Governments around the world have implemented a variety of incentive schemes to spur the adoption of residential PV, including tax credits, rebates, grants, net metering (i.e., feed-in tariffs) and renewable energy credit markets~\cite{Sarzynski2012, Bauner2015, Hagerman2016, Crago2017, Matisoff2017, Sunter:19, Boccard2021, Peasco2021, Kearns2022, Crago:23}.
Tax credits are a predominant incentive in the United States, where the federal government offers a 30\% tax credit on all community or rooftop solar installations~\cite{EnergyGov:2023:ITC}. 
Where applicable, state and local governments can also offer incentives, though these vary widely~\cite{Sunter:19} -- e.g., New York offers a 25\% state income tax credit and a property tax exemption on top of the existing federal incentives~\cite{NYSERDA:24}.
Interestingly, in their review of state and utility incentives for residential PV in the US, \citet{Matisoff2017} find that point of sale rebates (i.e., discounts applied at the time of purchase) are up to $8\times$ more effective in spurring adoption compared to tax credits worth the same amount.

\noindent\textbf{Inequities. \ }
Prior work has explored the distribution of rooftop PV installations across two primary demographic splits:

\noindent \textit{\textbf{Race and ethnicity}. \ } Multiple works have found that census tracts within the United States with Hispanic or Black population majorities had a deficit in the number of rooftop PV installations compared to the national average. \citet{Sunter:19} reported deficits of 61\% for Black majority tracts and 30\% for Hispanic majority tracts, while \citet{Dokshin2023RevisedEO} report deficits of 16\% for Black majority and 18\% for Hispanic majority tracts using different methodology.
Furthermore, \citet{Crago:23} found racial disparities in the existing financial returns of existing residential PV systems, due to different ownership and/or leasing models.

% \citet{Sunter:19} and \citet{Crago:23} consider the distributional impacts of solar in terms of adoption and financial returns, respectively, finding racial and income disparities.
 
\noindent \textit{\textbf{Income.}} Due to the capital expenditure required to install a PV system, household income is unsurprisingly %throughout the related literature as being strongly 
correlated with %the number of rooftop solar panels a household installs
rooftop PV installations~\cite{OShaughnessy:20}. In particular, in the United States, it has been found that households with a reported income greater than or equal to \$100k/year represent over half of all solar installations, while making up less than 25\% of the population in 2018. In a similar vein, 90\% of solar installations are owned by households with Prime ($680 - 740$) or Super-Prime ($>740$) credit scores~\cite{Barbose2020IncomeTA}. It has also been shown, however, that this trend is decreasing over time as more low- and middle-income households are installing rooftop photovoltaics. As of 2016 the median income of photovoltaic adopters has dropped to \$87,000 from \$100,000 in 2010 \cite{REAMES2020101612}. 

\section{Carbon Offset Potential and Siting Analysis}
\label{sec:analysis}

% \item Carbon Offset Potential and Siting Analysis (2.5)
% \begin{enumerate}
%     \item Problem Statement and Methodology
%     \begin{enumerate}
%         \item Problem Statement
%         \begin{itemize}
%             \item Results \/ Demographic Comparison \/ demo diff, both plots -- the bar plots of demographics against Carbon Offset Per Panel and Realized Potential (current figure 2)
%             \item results\/Carbon Offset\/CO vs Existing  2.png -- Plot showing that upper quartile of carbon offsetters have disproportionately few panels. (current figure 4)
%         \end{itemize}
%         \item Data sets
%     \end{enumerate}

% \rev{In this section, we start by analyzing the status quo of installations across the United States using several data sets. } \mo{sounds irrelevant and incomplete.}

In this section, we start with preliminaries about the measurements and data that we consider in the rest of the paper
before analyzing the socioeconomic and political attributes that characterize existing solar PV installations throughout the United States.

\vspace{-0.2cm}
\subsection{Problem and Methodology}
\label{sec:prob}
In conducting this analysis, we aim to answer the following questions that consider possible inequities, energy inefficiencies, and carbon inefficiencies amongst existing solar PV installations:
% To understand the inequity and reveal the carbon inefficiency, we study the annual carbon offset potential, annual solar energy generation potential, and demographical distribution for locations with existing solar PV installations. 

\vspace{-0.1cm}
\begin{enumerate}
    \item How do different locations in the United States differ in their energy generation and carbon offset potential? 
    \item Are the existing solar PV installations energy-efficient or carbon-efficient?
    \item Are the existing solar PV installations equitable across racial, economic, and political demographics? 
\end{enumerate}

To answer these questions, we outline some key definitions that contextualize the necessary information about each installation:

% \noindent\textbf{Definitions. \ }
% We first define the solar energy generation potential, carbon offset potential, and realized potential. 
% We also outline the three demographic aspects we consider. 
% \mo{make sure paragraph titles are consistent.}
\begin{enumerate}[leftmargin=*, itemsep=0.1cm]
    \item \textbf{Solar Energy Generation Potential} is the amount of sunlight energy that can be harvested by installing rooftop solar panels in a given location. It depends on the geographical location, local weather characteristics, a rooftop's physical characteristics, and the energy efficiency of the solar panels~\cite{Bashir:2019:SolarTK}. It is measured in kilowatt-hours (kWh); a higher value is better. 
    \item {\color{blue} \textbf{Carbon Offset Potential} is the amount of avoided carbon emissions by using electricity from the rooftop solar instead of the electricity from the local electric grid. It depends on the solar energy generation potential and the carbon intensity of the local electric grid as well as the timing of local energy demand in relation to viable solar generation hours.
    Recall that carbon intensity quantifies the amount of CO$_2$ emissions per unit of electricity generated. It is a function of the energy sources that supply grid electricity in a particular location. Carbon offset potential is measured in kilograms of carbon emissions, and a higher value is better.}
    \item \textbf{Realized Potential} is the number of existing solar panel installations as a percentage of the potential solar PV installations. Its value can be between 0 and 100; a higher value indicates low potential for new installations and vice versa.
\end{enumerate}

To analyze inequity across demographics, we look at the distribution of different parameters (e.g., installations, carbon offset) across different demographic groups, which are detailed below in our dataset descriptions.

\subsection{Datasets}
\label{sec:data sets}
In this section, we detail the datasets used in our subsequent analysis.
% We first outline key definitions and then outline the data sets we use in answering our questions. 
\begin{table}[t]
    \centering
    \footnotesize
    \caption{Description of datasets used in our analysis.}
    \vspace{-0.3cm}
    \label{tab:data sets}
     \begin{tabular}{|l|c|c|}
         \hline
          \textbf{Metric} & \textbf{Dataset} & \textbf{Time Span}\\
         \hline
         \hline
         Solar energy generation potential & Google's & 8/17/2015\\ \cline{1-1}
         Carbon offset potential & Project & to\\ \cline{1-1}
         Existing solar installation count  & Sunroof & present
         % \tablefootnote{Data retrieved in May 2024.} 
         \\ \hline \hline
         Demographic information & American Community  & 2016 to \\ 
         (racial, economic) & Survey Dataset  & 2020 \\ \hline
         Demographic information (political) & MEDSL Election Voting & 2020\\
         \hline
    \end{tabular}
    \vspace{-0.45cm}
\end{table}
% source for sunroof: https://green.googleblog.com/2015/08/project-sunroof-mapping-planets-solar.html
% \subsubsection{\textbf{Data sets}}
As outlined in \autoref{tab:data sets}, we use four datasets in our analysis to gather various information: \textbf{(1)} Google's Project Sunroof dataset~\cite{sunroof} retrieved in 2024, \textbf{(2)} the U.S. Census' 2016-2020 American Community Survey (ACS5)~\cite{ACS}, \textbf{(3)} Ember's annual energy generation dataset~\cite{Ember:22}, and \textbf{(4)} a political voting dataset compiled by the MIT Election Data + Science Lab (MEDSL)~\cite{MEDSL:20}. The Project Sunroof and ACS5 datasets are collected at a ZIP code granularity, while the energy generation and political voting datasets are available at a state level. %\mo{what about political vote data?}
Next, we describe each dataset and how we augment them to handle their drawbacks. In \autoref{sec:setup-toolkit}, we highlight our contribution in releasing the \toolkit toolkit for public use.

% \smallskip
\vspace{-0.1cm}
\subsubsection{\textbf{Project Sunroof Dataset}}
\label{subsec:sunroof}
The Project Sunroof dataset is compiled from a combination of machine-learned computer vision (CV) and human labeling of buildings, existing solar panels, and viable solar panel locations in Google satellite images of the US. 
These measurements are combined with the yearly average sunlight in each ZIP code to calculate the region's potential solar energy generation down to a single solar panel on a given rooftop.  

Sunroof uses a CV model to estimate the topology of each building's roof and shading profile in estimating \emph{energy generation potential}, considering the sun's angle and surrounding trees.
{\color{blue}The dataset also includes an estimated value of each ZIP code's \emph{carbon offset potential}. The calculation of which uses the energy generation potential to estimate the \emph{carbon offset potential} of solar generation using subregion CO2-equivalent non-baseload emission rates from the eGRID dataset from the U.S. Environmental Protection Agency (EPA)~\cite{epa_egrid_2015}. This calculation is described in the Project Sunroof methodology~\cite{sunroof}.}
Finally, Sunroof's CV model labels regions of each roof according to their likelihood of being an existing solar PV installation -- humans then label high-likelihood regions as an existing install or not.

We use energy generation potential, carbon offset potential, and existing install count values from the dataset. 
Since Sunroof only provides the existing install count for a given ZIP code, we need to estimate the potential installation count while accounting for the differences across various ZIP codes. 
To do that, we scale each value using the \textit{percent-covered} value given in the dataset. For instance, if 50\% of a ZIP code is covered, we multiply the number of existing installations by 2 to estimate the total for a given ZIP code. 

% From the Project Sunroof data set, we use two primary values:
% \begin{itemize}
%     \item \textbf{Existing install count :} The Sunroof computer vision model first labels regions of each roof according to their likelihood of being an existing solar PV installation -- high likelihood regions are then labeled as such after confirmation by a human.
%     \item \textbf{Carbon offset (metric tons):} This is an estimation of the amount of  CO$_2$ emissions that would be prevented (i.e., offset) if all of the viable rooftops in a given ZIP code contained solar installations. This quantity heavily depends on the existing energy sources on the area's electric grid.
% \end{itemize}

We note four additional potential drawbacks of this dataset: \textbf{(1)} The dataset's coverage within the continental United States is incomplete. 
Only 11,516 out of 33,774 ZIP codes have coverage, and of those covered, very few have full coverage.  
Furthermore, those zip codes with coverage are naturally skewed towards regions with more homes, thus excluding many rural areas. 
\textbf{(2)} There is a heavy reliance on machine-learned computer vision models, which are necessary for the sheer scale of the data but are imperfect and provide no guarantees on the accuracy (none of which are stated by Google). {\color{blue}
\textbf{(3)} The dataset's estimates assume uniformity in the size and efficiency of the solar panels as well as necessary grid infrastructure. \textbf{(4)} The estimation of carbon offset potential does not account for the timing of energy generation and demand, which are necessary for precise calculations of carbon offset potential. This also does not take into account the well known phenomenon of misaligned PV generation and demand, deemed the "duck curve" problem~\cite{Calero2022DuckCurveMI}.

Since there is a lack of suitable data to address the aforementioned issues, we only note that the availability of such data would further improve the accuracy of the method we outline in the rest of the paper.  Full documentation on the Project Sunroof methodology is publicly detailed by Google~\cite{sunroof}.}

% All other issues must be accepted however as there is no better data set available.
% \smallskip
\vspace{-0.1cm}
\subsubsection{\textbf{American Community Survey Dataset}}
The American Community Survey is a monthly / yearly survey given to randomly selected households across the country that is used to update a running estimate maintained by the US Census Bureau.  
Participants share their income, race, employment, and other demographic and economic information, compiled at the ZIP code's granularity and used as a basis for policy development alongside the Census.

\begin{figure}[t]
    \vspace{-1em}
    \centering    
    \includegraphics[width=\columnwidth]{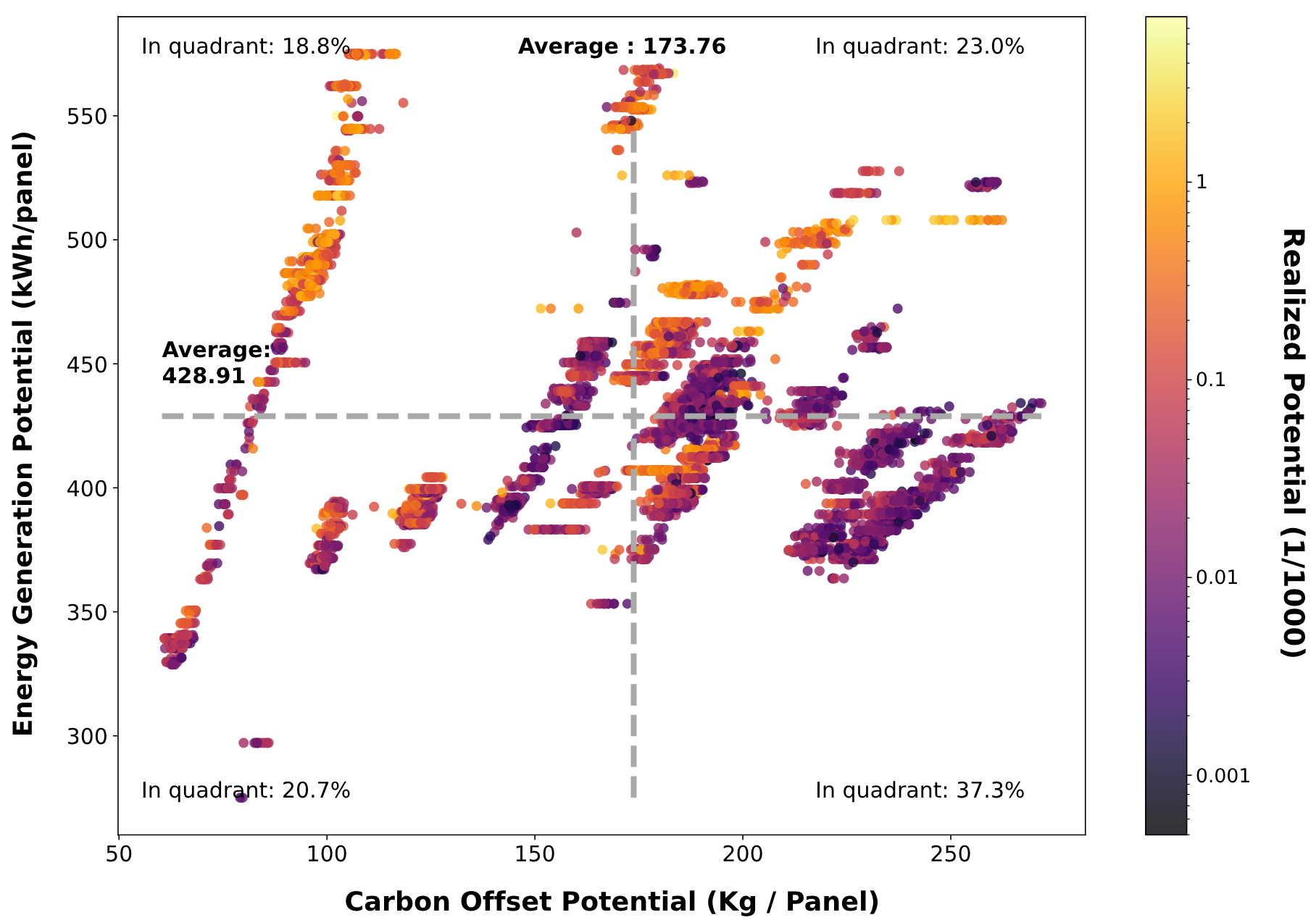}
    \vspace{-1em}
    \caption{Solar energy generation potential (y-axis) and carbon offset potential (x-axis) per 400W panel. Each circle represents a single ZIP code, and the circle's color represents the magnitude of realized potential in that ZIP code.}
    \label{fig:carbon-energy-potential}
    \vspace{0.5cm}
    \centering
    \includegraphics[width=\linewidth]{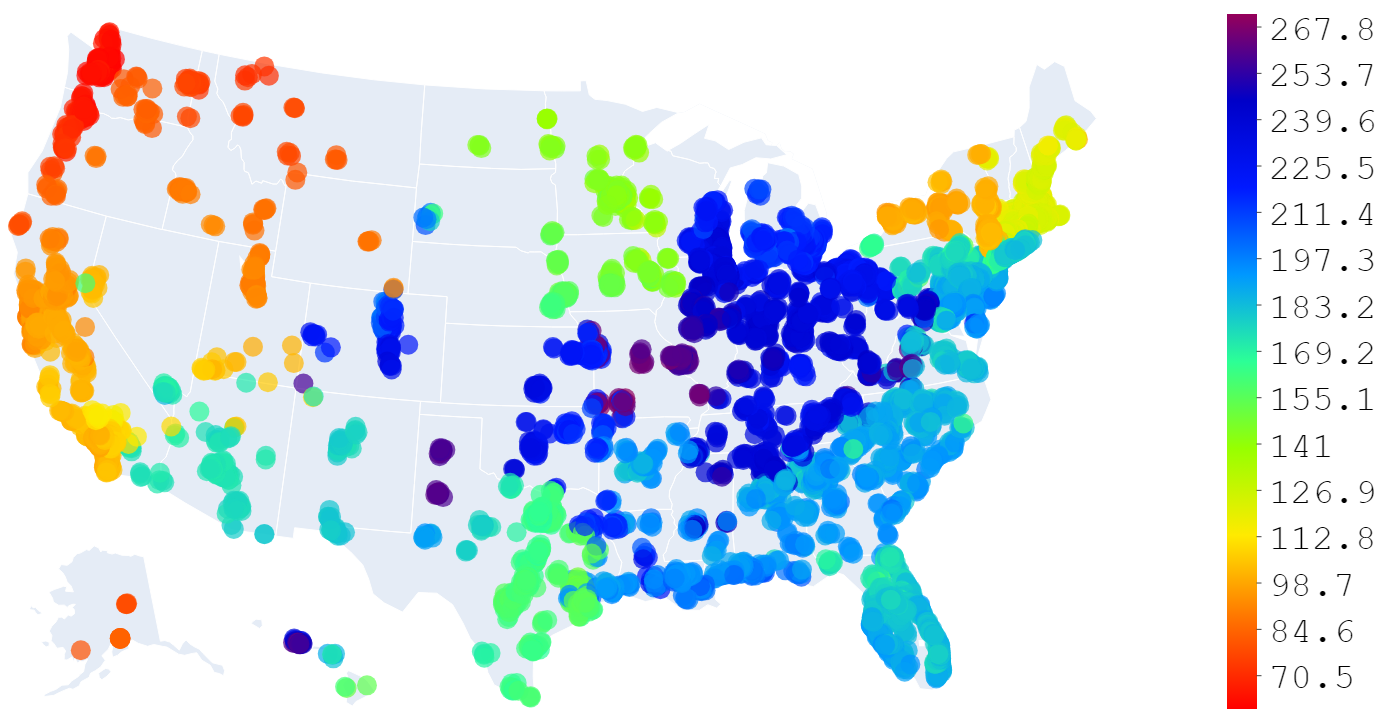} \vspace{-0.5em}
    \caption{Map of {\color{blue}estimated} carbon offset potential (kg/panel) for each ZIP code represented in the dataset.}
    \label{fig:co_per_cap_map}
    \vspace{-0.5cm}
\end{figure}

\subsection{Analysis}
\label{Analysis}

\begin{figure*}[t]
    \vspace{-1em}
    \centering
\subfloat[Realized Potential ($\times$ National Average)]{
	\begin{minipage}[c][0.5\width]{
	   0.5\textwidth}
	   \centering
	   \includegraphics[width=\textwidth]{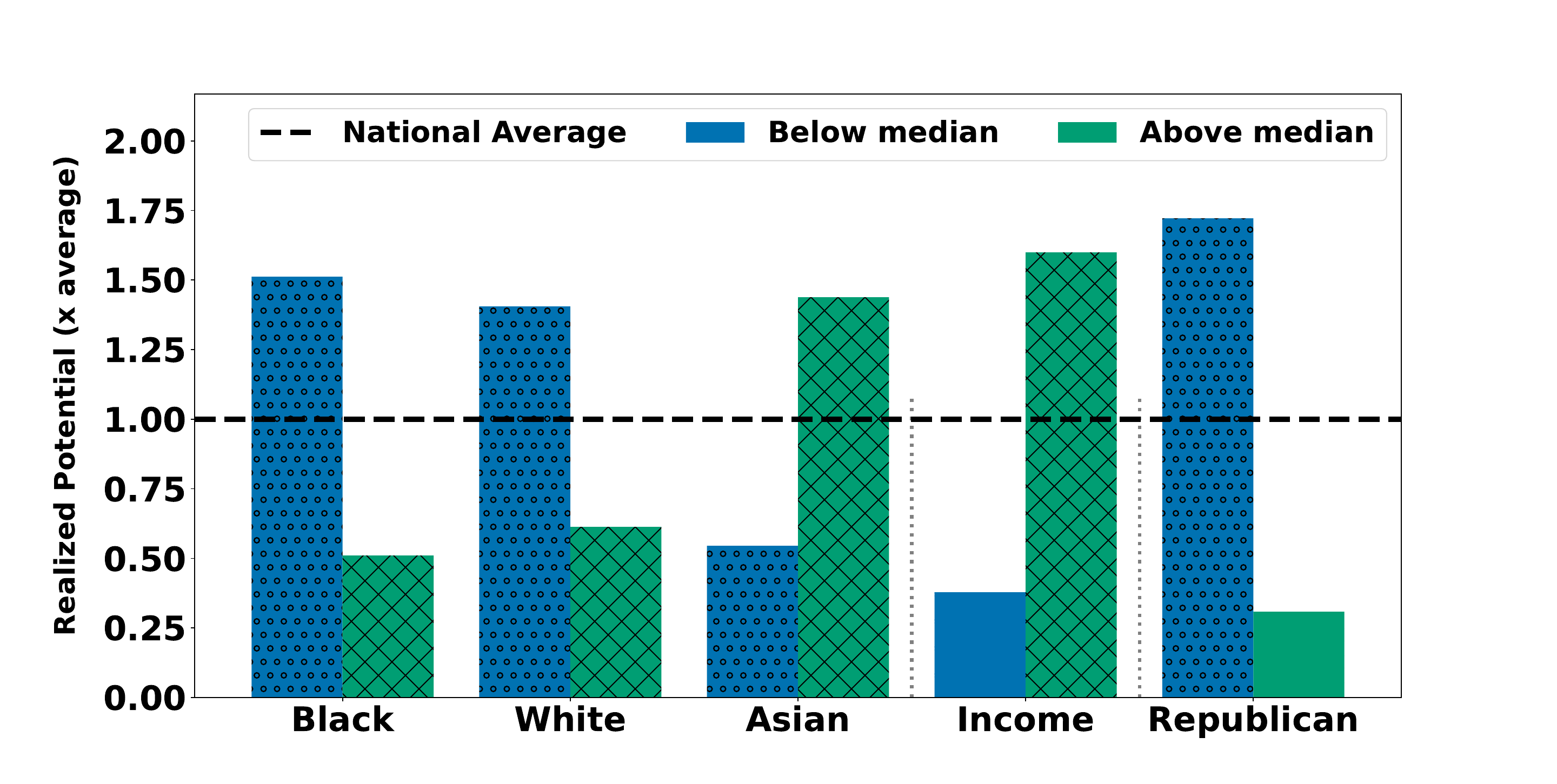}
    \vspace{-0.3cm}
	\end{minipage}}
  \subfloat[Carbon Offset Potential ($\times$ National Average)]{
	\begin{minipage}[c][0.5\width]{
	   0.5\textwidth}
	   \centering
	   \includegraphics[width=\textwidth]{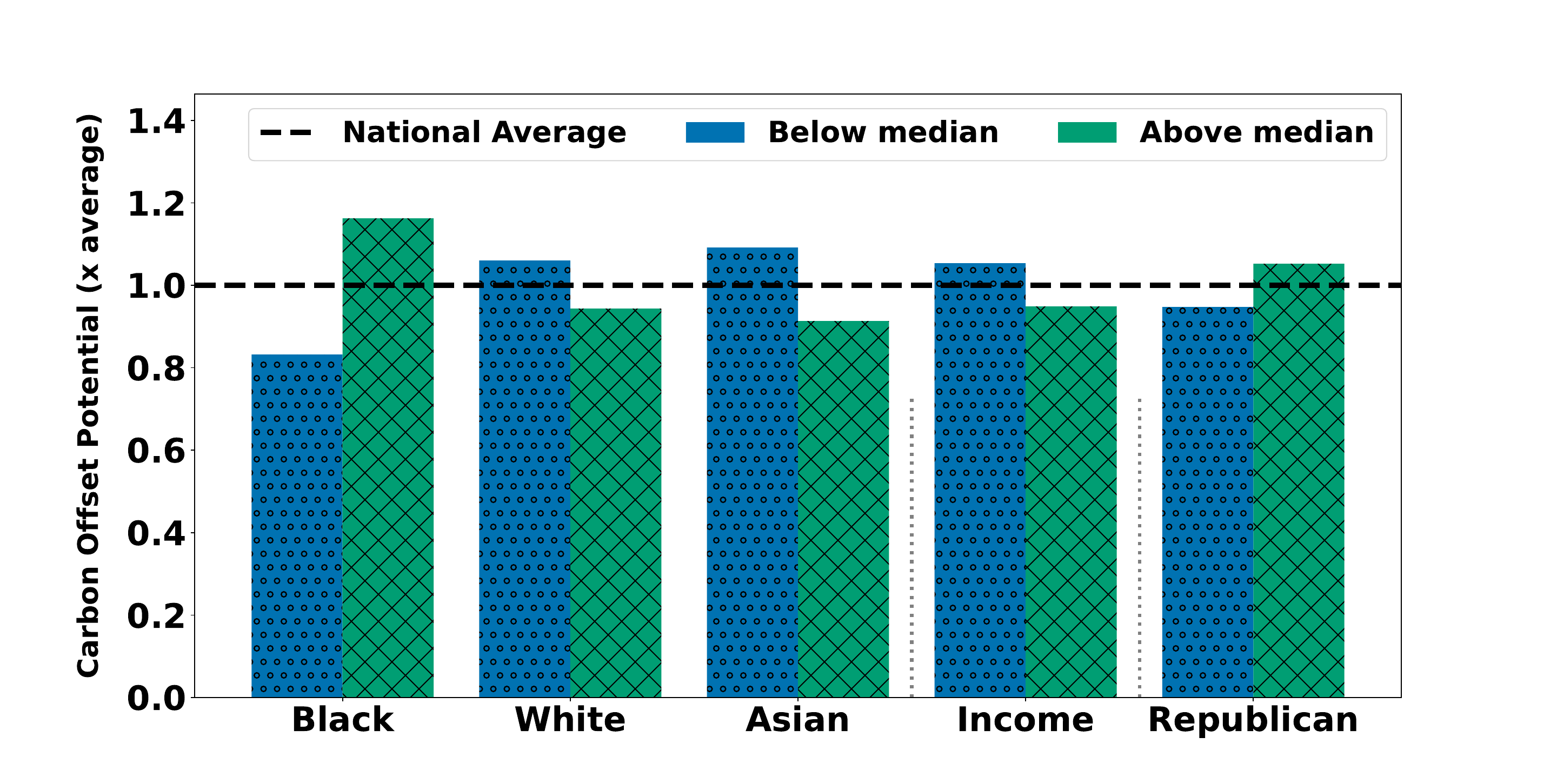}
     \vspace{-0.3cm}
	\end{minipage}
 }
    \vspace{-0.3cm}
    \caption{Inequity and inefficiency analysis using (a) the realized potential (existing installs / possible installs) for a given group compared to the national average and (b) the carbon offset potential per panel for a given group compared to the national average. A value of 1 means that the given group has the same statistics as the national average. Each group on the x-axis accounts for all the states. Green bars with a cross pattern represent the states with above median values, and blue bars with circles represent those with below median values. State-level statistics are computed by aggregating ZIP code-level data for all groups except political inclination, which is already at the state level. 
    % Average carbon offset per panel and realized potential (existing installs / possible installs) compared to the national average (proportional difference) for states with above and below the national median of different demographics.
    }
    \label{fig:inequity-analysis}
    \vspace{-0.4cm}
\end{figure*}

In what follows, we use the aggregation of this data across the 5-year period 2016-2020 (henceforth abbreviated ACS5). We use the estimated values calculated for the dataset while removing all ZIP codes with too few samples to calculate reasonable estimations, which are labeled appropriately.

From the 2016-2020 ACS5, we use the following data points:
\begin{itemize}[leftmargin=*, topsep=0.1cm]
    \item \textbf{Median income:} Median Yearly Income (USD) estimated at a ZIP code granularity.
    \item \textbf{Total population:} Estimation of the total population living in a particular ZIP code.
    \item \textbf{Racial demographic composition:} Estimations of the (fraction of) population in a given ZIP code that self-identifies as belonging to a particular race or ethnicity. We disaggregate our analysis according to the White, Black, Hispanic, and Asian racial groups. 
\end{itemize}

We chose the 2016-2020 5-year window as it was the first to implement more accurate estimation techniques~\cite{censusCommitmentQuality} and is least-impacted by a non-response bias identified by the Census Bureau during the years of the COVID-19 pandemic~\cite{census_multiyear_acs_2020}.

% \smallskip
\vspace{-0.15cm}
\subsubsection{\textbf{Energy Generation Dataset}} 
We use a summary of yearly energy data compiled by Ember~\cite{Ember:22}. This data aggregates information from authoritative sources including the Energy Institute (EI) and U.S. Energy Information Administration (EIA).  We use energy generation and emissions estimates by fuel type (e.g., solar, wind, coal, gas, etc.) at a state level for all the states and total across the US.
We use the 2023 compilation of Ember's data as it is the most recent release available at the time of writing~\cite{Ember:22}. 
% \adaml{2023 data is released as of August it seems -- we should update our results}

% \smallskip
\vspace{-0.15cm}
\subsubsection{\textbf{Political Voting Dataset}}
We use a state-level election result dataset compiled by the MIT Election Data + Science Lab (MEDSL)~\cite{MEDSL:20}. In our analysis, we use the results of the 2020 presidential election, notably the votes cast for the Democratic (Joseph Biden) and Republican (Donald Trump) candidates.  We use this data because it is the most accurate summary of results in the 2020 election -- we note that the data does not include write-in votes.

\vspace{-0.1cm}
\subsubsection{\textbf{Combining Datasets}}
% \label{data set combination}
In our analysis, we interchangeably use ZIP code-level and state-level granularity values. For ZIP code-granular experiments, we use primarily the Project Sunroof and ACS5 datasets, and of those, we use ZIP codes that are \textit{both} covered by the Project Sunroof dataset and considered to have a \textit{large enough sample size} in the ACS5. This gives 10,559 ZIP codes in the combined set. In some experiments, outliers are omitted for better visualization and are explicitly mentioned where applicable.

In the state-level granularity analysis, we aggregate our ZIP code-level datasets at the state-level while omitting US territories and the District of Columbia. For the Project Sunroof and ACS5 data, we average over the 10,559 ZIP codes of the ZIP code-level dataset and disaggregate by state to maintain consistency with the ZIP-code granularity analysis.
Next, we present our analysis of the datasets to answer the three questions posed at the start.

%     \item Solar Potential Analysis
%         \begin{itemize}
%             \item Sunlight vs Carbon Offset figure which was requested (in carbon offset folder, not current figure) 
%             \item Energy Gen breakdown by State (current figure 5)
%             \item We can replace this with all states instead of exemplar if we are removing that. 
%         \end{itemize}
%     \item Solar Siting Analysis
%         \begin{itemize}
%             \item Map of Carbon Offset by Zip (Figure 7)
%             \item 2x3 maps (Figure 8)
%         \end{itemize}
% \end{enumerate}

% \mo{make subsubsection consistent between 3.2 and 3.3.}
\vspace{-0.1cm}
\subsubsection{\textbf{Energy and Carbon Offset Potential}}
We first examine the solar energy potential and the carbon offset potential across all the available ZIP codes.
As shown in \autoref{fig:carbon-energy-potential}, along the $y$-axis, there is a significant difference in energy generation potential across the ZIP code with values ranging from 250 kWh to almost 600 kWh. 
As outlined in \autoref{sec:background}, this difference is not only due to variations in the amount of sunlight each ZIP code receives but also due to the unique characteristics of homes in a given ZIP code.
The coefficient of variation (CoV) -- calculated as the standard deviation over mean -- for the solar energy generation potential is 0.12, quantitatively depicting the widespread of generation potential across sites. 
Similarly, along the $x$-axis in~\autoref{fig:carbon-energy-potential}, we can observe a wide range of carbon offset potential distribution. Some regions have a carbon offset potential of less than 50 Kg of carbon dioxide equivalent per panel to more than 250 Kg per panel. 
It has a higher value of CoV, 0.28, than the solar energy potential because the carbon offset potential is influenced by changes in both energy generation potential and the carbon intensity of the electric grid in the region, which varies significantly across locations.  In \autoref{fig:co_per_cap_map}, 
we visualize the \textit{carbon offset potential} at a ZIP code granularity.
We also observe that the energy generation and carbon offset potential are not well correlated. 
Pearson's correlation coefficient is 0.196; the energy generation potential impacts the overall carbon offset potential, but the region's carbon intensity effect dominates.

\vspace{0.1cm}
\noindent
\emph{\textbf{Key Takeaway.} Yes, the solar energy generation potential and carbon offset potential vary significantly across locations. Also, the locations with the highest energy generation potential do not always have the highest carbon offset potential, stressing the need for a strategic approach to future rooftop solar capacity growth.
}

\vspace{-0.15cm}
\subsubsection{\textbf{Energy and Carbon Efficiency of Existing Installations}} 
In \autoref{fig:carbon-energy-potential}, the color gradient indicates the number of installations, with lighter-colored dots representing areas with many installations. 
It reveals a significant concentration of existing solar installations in regions with higher-than-average energy generation potential but lower-than-average carbon offset potential regions. 
The top-left quadrant, which corresponds to regions with high energy generation potential but low carbon offset potential, has many lighter-colored dots. This suggests that many solar installations are located in areas that generate significant energy but yield relatively low carbon offsets, leading to prevalent carbon inefficiency in the current strategy of deploying solar panels.

\begin{table}[t]
    \centering
    \footnotesize
    \caption{ZIP code level statistics of the data in \autoref{fig:inequity-analysis} for three racial demographics and one income group.}
    \vspace{-0.3cm}
     \begin{tabular}{ |l|c|c|  }
         % \hline
         % \multicolumn{3}{|c|}{Underutilized demographics at ZIP code granularity} \\
         \hline
          \textbf{Demographic} & \textbf{Carbon Offset Potential} & \textbf{Realized Potential}\\
          & \textbf{(\% Average)} & \textbf{(\% Average)} \\
         \hline
         Black pop. (> median)  & +6.9 \% & -34.0 \%\\
         Asian pop. (> median) & -5.6 \% & +43.7 \%\\
         White pop. (> median) & +1.5 \% & +6.0 \%\\ \hline
         Median Income (< median) & +6.8 \%  & -29.4 \% \\
         
         \hline
    \end{tabular}
    \label{tab:demo_diff_zip}
    \vspace{-0.5cm}
\end{table}

\begin{figure*}[t]
    \centering
  \subfloat[Yearly average sunlight]{
	\begin{minipage}[c][0.55\width]{
	   0.33\textwidth}
	   \centering
	   \includegraphics[width=1\textwidth]{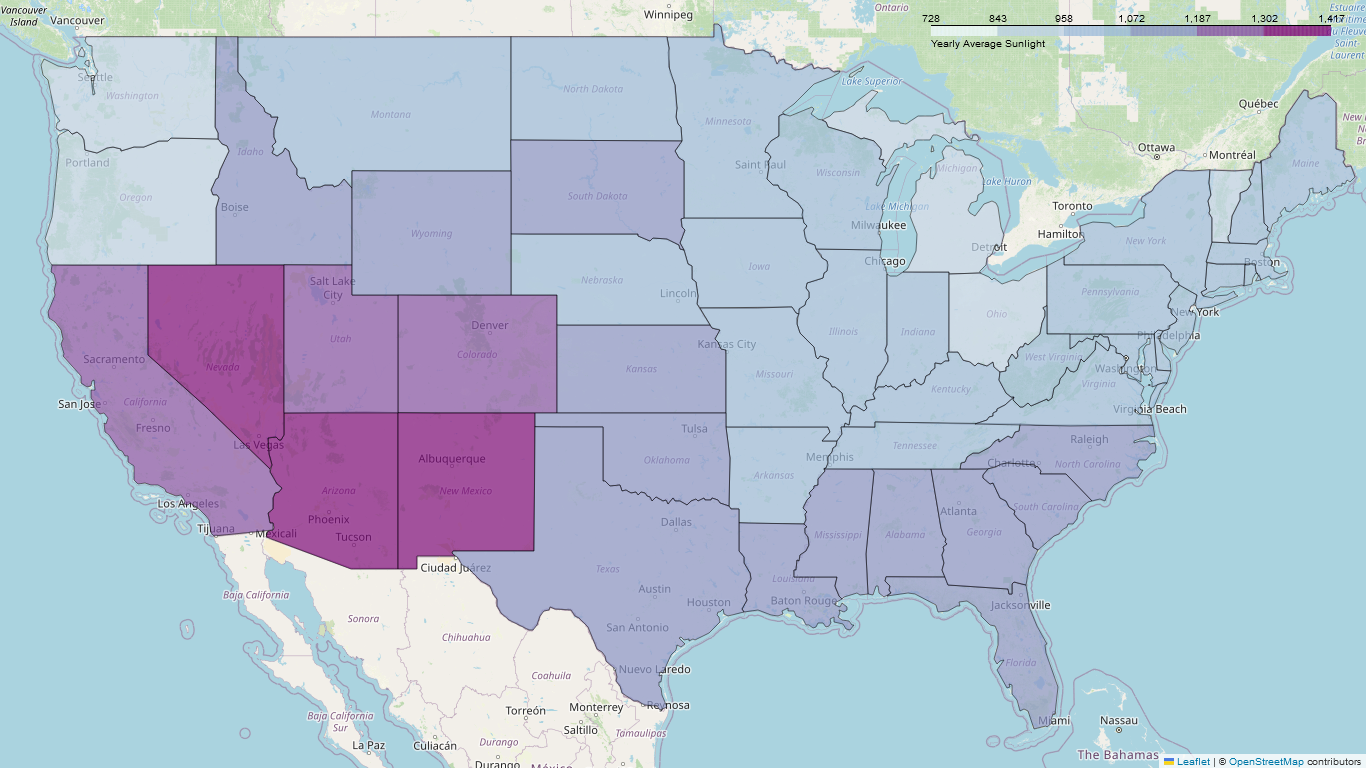}
	\end{minipage}}
   \subfloat[Black population proportion]{
	\begin{minipage}[c][0.55\width]{
	   0.33\textwidth}
	   \centering
	   \includegraphics[width=1\textwidth]{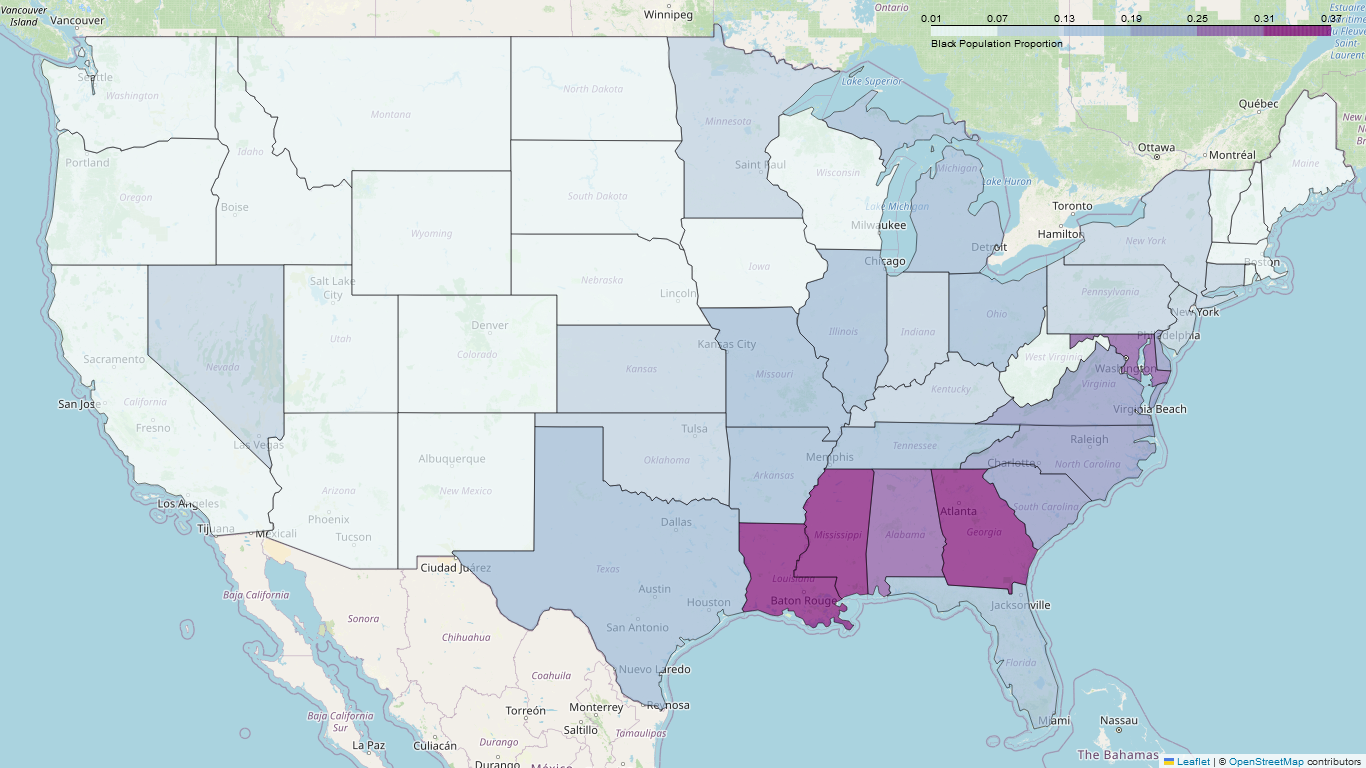}
	\end{minipage}}
   \subfloat[Median income]{
	\begin{minipage}[c][0.55\width]{
	   0.33\textwidth}
	   \centering
	   \includegraphics[width=1\textwidth]{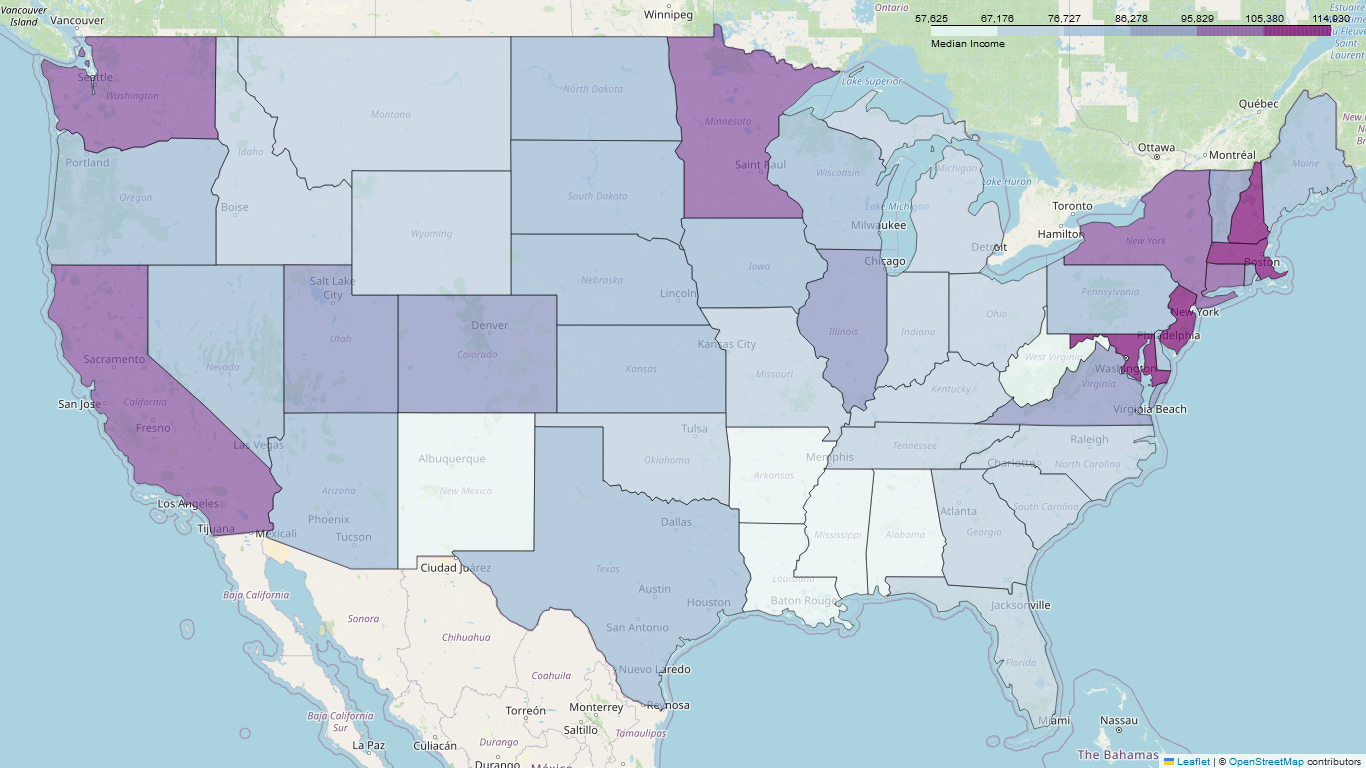}
	\end{minipage}}
    \hfill
      \subfloat[Republican voter proportion]{
	\begin{minipage}[c][0.55\width]{
	   0.33\textwidth}
	   \centering
	   \includegraphics[width=1\textwidth]{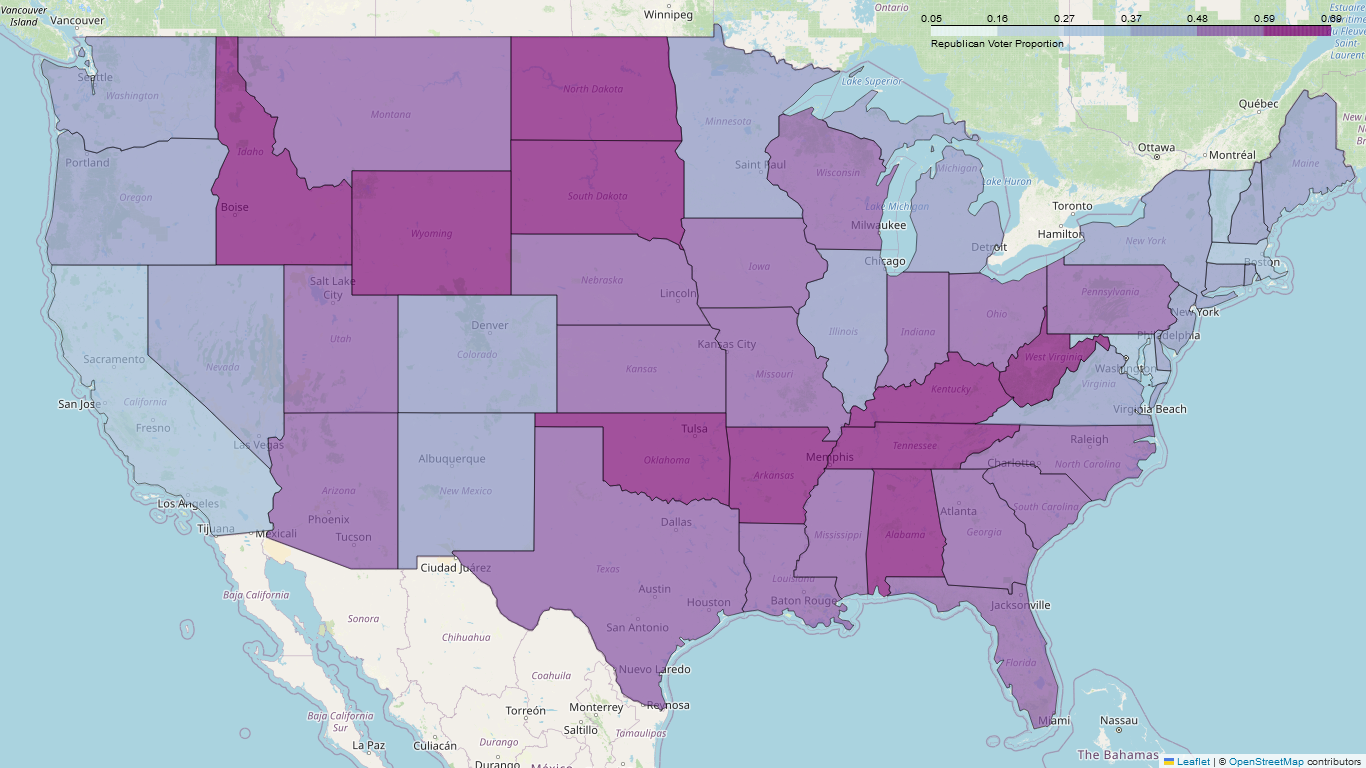}
	\end{minipage}}
       \subfloat[Estimated carbon offset per panel]{
	\begin{minipage}[c][0.55\width]{
	   0.33\textwidth}
	   \centering
	   \includegraphics[width=1\textwidth]{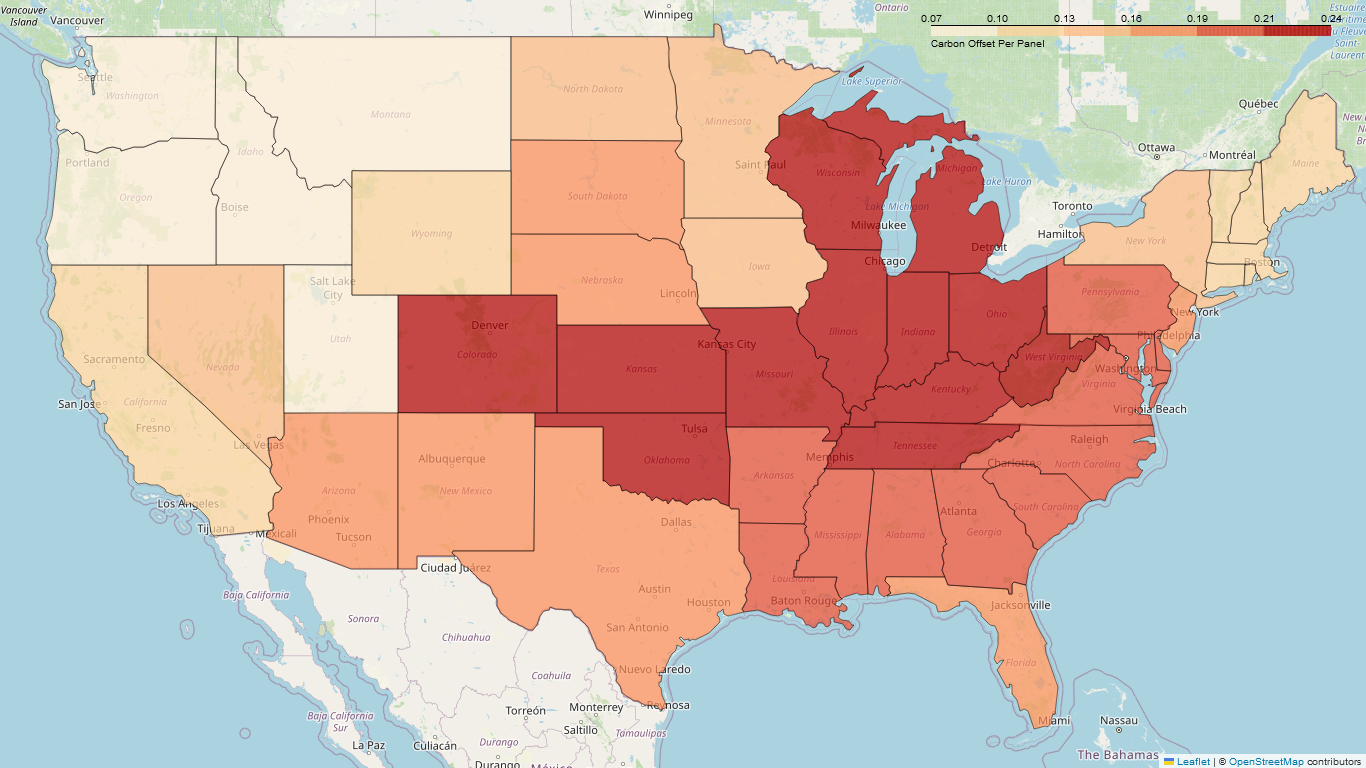}
	\end{minipage}}
       \subfloat[Realized potential]{
	\begin{minipage}[c][0.55\width]{
	   0.33\textwidth}
	   \centering
	   \includegraphics[width=1\textwidth]{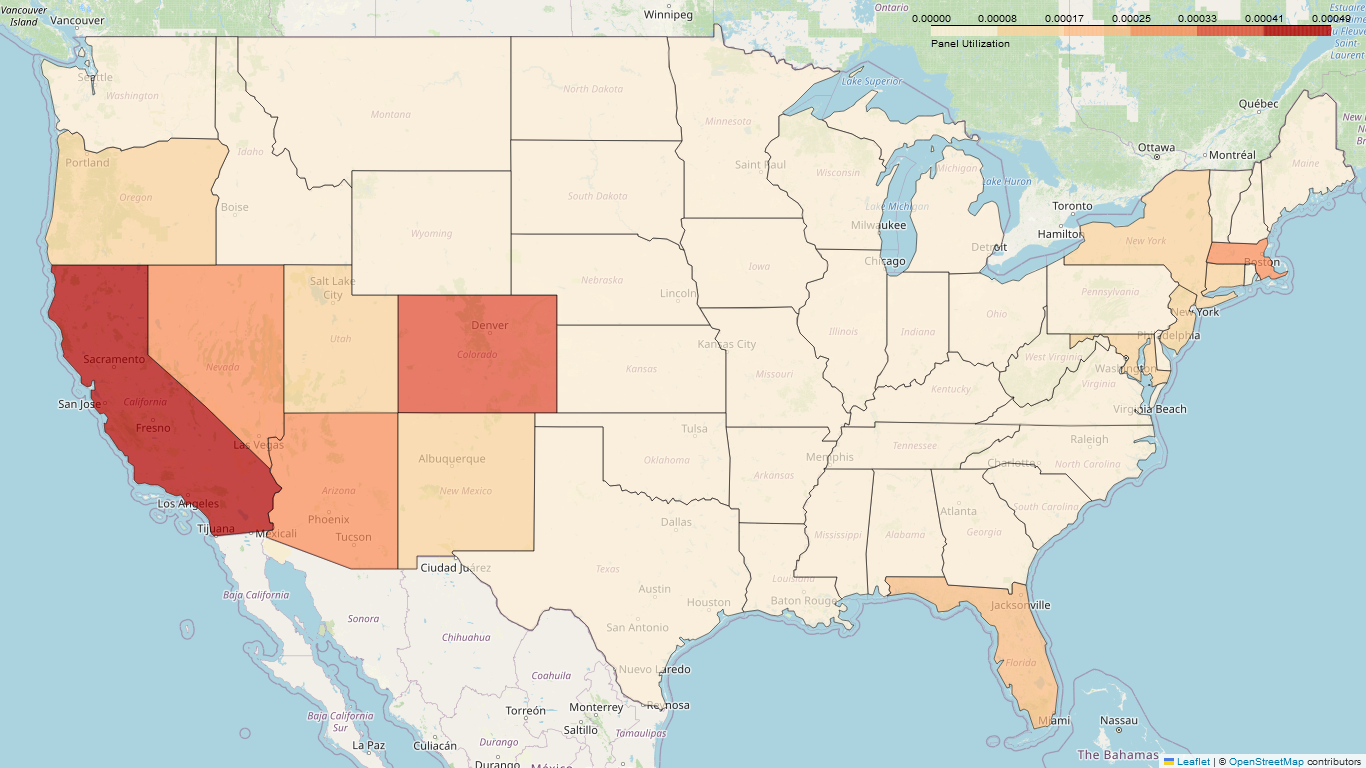}
	\end{minipage}}\vspace{-1em}
    \caption{State-level granularity maps of relevant features.}
    \vspace{-0.5cm}
    \label{fig:state maps}
\end{figure*}

The top-right quadrant of the figure, characterized by high energy generation potential and high carbon offset potential, is populated by lighter-colored dots, indicating a high concentration of solar installations. This is the ideal scenario for solar deployment since these regions are both energy-efficient and carbon-efficient.  The high density of installations in this quadrant suggests that some progress has been made in placing solar panels where they can have the greatest dual impact.
Unfortunately, only 23\% of all locations fall in this quadrant, limiting the potential. 
Conversely, the darker-colored dots in the figure, representing regions with fewer installations, are predominantly found in areas with higher carbon offset potential, such as the bottom-right quadrant, which contains 37\% of all the sites. 
While these regions may generate less energy, they offer substantial carbon reduction benefits, making them valuable targets for future solar deployment. 

\vspace{0.1cm}
\noindent
\emph{\textbf{Key Takeaway.} The existing solar installations are understandably located in higher-energy generation potential regions (top half of the graphs), which accounts for 42\% of all the locations. Unfortunately, the majority of the sites with high carbon offset potential have low energy generation potential (bottom right quadrant) and, therefore, have been neglected by the current installation trends.
}

\vspace{-0.1cm}
\subsubsection{\textbf{Socioeconomic Inequities in Existing Installations}} 

\autoref{fig:inequity-analysis} highlights significant inequities and inefficiencies in the existing solar installations across different demographic, income, and political affiliation groups. 
As shown in \autoref{fig:inequity-analysis}(a), the realized potential for the above-median Black population and below-median income groups is consistently lower than the national average. 
However, in \autoref{fig:inequity-analysis}(b), we observe that states with an above-median Black population and below-median income have a higher carbon offset potential than the national average.
This outcome is consistent with prior findings demonstrating inequity towards black populations and low-income groups but additionally highlights the hidden carbon cost of inequity in solar installations.

The analysis for the white and Asian racial groups has the opposite trend. 
Higher-than-median white population areas have low carbon offset potential and few existing solar installations.
However, the states with higher-than-median Asian populations have a higher number of existing installations than the national average but a lower-than-median carbon offset potential. 
It is worth noting that the disparity for all of these groups is above 35\%, showing a high inequity. 
We also analyze the inequity and inefficiency across demographics based on political inclination. 
We observe that states that primarily voted for Republicans have a higher-than-average carbon offset potential, by around 5\%, but almost 70\% less existing installations than the national average. 

Finally, in addition to the state-level statistics, we also analyze the underlying ZIP code-level statistics, as shown in \autoref{tab:demo_diff_zip}. 
We see a similar disparity at the ZIP code level, which shows that the problem of inequitable distribution across different demographics causing carbon inefficiency is not specific to a small number of regions within a given state. 
Instead, these inefficiencies are well spread across all the ZIP codes within the states. 

\vspace{0.1cm}
\noindent
\emph{\textbf{Key Takeaway.} The existing solar installations are neither inequitable nor carbon-efficient. Locations have lower-than-average existing installations if they have a higher-than-median Black population proportion, lower-than-median income, or higher-than-median Republican voting proportion. Unfortunately, these are also the locations that have higher-than-average carbon offset potential. 
% Therefore, the existing installations are neither equitable nor carbon-efficient.
}
% \mo{average vs. median? are these used intentionally here?}

% \adaml{everything demographic should be median, everything solar should be average, I can check for this throughout and change}

% \clearpage
\section{Solar Siting Strategies}
\label{sec:strategies}
In this section, motivated by the carbon offset and demographics observations of \autoref{sec:analysis}, we design and simulate \textit{siting strategies} to guide future rooftop PV installations.  
In \sref{Sections}{sec:setup-toolkit} and \ref{sec:evaluation}, we simulate the outcomes of these strategies and evaluate their impacts on energy generation, equity, and decarbonization.

% \subsection{The Status Quo}
\vspace{0.1cm}
\noindent\textbf{1 -- The Status Quo.}
To motivate the strategies we design below, we first review the current distribution of installations analyzed in the prior sections. 
In \autoref{fig:state maps}(f), we show a map of realized potential, which corresponds to the number of existing rooftop PV installations by state.  
As discussed above, these results show that certain states (e.g., California and Colorado) are responsible for a disproportionate fraction of the installed rooftop solar capacity in the United States.
To simulate the behavior of a ``business as usual'' solar siting strategy, we evaluate the impact of adding a certain number of panels $N$, weighted proportionally to the existing distribution of rooftop PV in the US.  
This means that areas with the bulk of existing installations will also proportionally receive the bulk of new installations in the future.  While this is not necessarily reflective of true future projections for rooftop PV in the United States (e.g., current projections show more ongoing growth in the American Midwest PV market \cite{SEIA_2024_major_projects}), it serves as a useful baseline to understand the implications of continued distributional impacts.

% \subsection{Optimizing for Energy Efficiency}

\vspace{0.1cm}
\noindent\textbf{2 -- Optimizing for Energy Efficiency.}
Given that the amount of viable sunlight for energy production varies greatly throughout the US, it is natural to consider a siting strategy to maximize the amount of electricity generated relative to the number of panels deployed.  To simulate the behavior of such a strategy, we evaluate the impact of adding $N$ panels in the order of energy potential that includes average sunlight and other factors as explained in \autoref{sec:prob} -- i.e., panels are installed in the most energy efficient ZIP code of the country until the available rooftop space is exhausted, followed by the second most efficient ZIP code, and so on.

% \subsection{Optimizing for Equity}

\vspace{0.1cm}
\noindent\textbf{3 -- Optimizing for Equity.}
Based on the analysis in \autoref{sec:analysis} and prior work, it is equally intuitive to consider a siting strategy that deploys PV panels to alleviate existing socioeconomic inequities.  We simulate the behavior of two such strategies, where each adds $N$ panels nationwide in descending order of Black population proportion and ascending order of low median income, respectively.  These strategies target populations identified in \autoref{sec:analysis} as having a combination of low realized potential and high carbon offset.

% \subsection{Optimizing for Carbon Efficiency}
\vspace{0.1cm}
\noindent\textbf{4 -- Optimizing for Carbon Efficiency.}
Since each additional PV panel has a different carbon offset (the amount of CO$_2$ avoided due to its installation) based on its location, it is also natural to consider a siting strategy that deploys PV panels in locations that currently have very high carbon-emitting grids.  
To simulate the behavior of such a strategy, we evaluate how adding $N$ panels in descending order of carbon offset per panel -- i.e., panels are installed in the ZIP code with the highest carbon offset in the country until available rooftop space is exhausted, followed by the second highest carbon offset ZIP code, and so on.

% \subsection{Equity-Aware Decarbonization}
\vspace{0.1cm}
\noindent\textbf{5 -- Multi-Objective Decarbonization.}
{\color{blue} The previous siting strategies optimize for \textit{one} specific attribute (e.g., energy efficiency, equity, carbon offset). However, in a realistic strategy, it is crucial to consider a more holistic strategy that balances all objectives: energy efficiency, carbon efficiency, and equity in various terms, e.g., race and income.  
To simulate the behavior of this ``balancing'' strategy, we add $N$ panels in a ``round robin'' order -- i.e., we choose the first ZIP codes in each sorted list (energy, carbon, equity) before cycling through to the second ZIP code in each, and so on.  This strategy effectively gives ``equal weight'' to each objective term while placing installations. In our simulations, this is the only strategy that addresses the multi-objective optimization problem of solar siting.}

\begin{comment}
\adaml{I think this isn't necessary here}
In this section, we present specific experiments and results, seeking to answer the following three questions, motivated by disparities identified in the introduction.
\begin{enumerate}
    \item \textit{Which demographics are most correlated with areas that exhibit high carbon offset potential, but few current installations?}  
    \item \textit{Can we identify potential mechanisms that may cause these demographics to be correlated with this issue?}
    \item \textit{Out of the nationwide data, can we highlight ZIP codes and states that exemplify this issue and would benefit from targeted interventions?}
\end{enumerate}
\end{comment}

\section{Experimental Setup and Toolkit}
\label{sec:setup-toolkit}
In this section, we describe our experimental setup to evaluate the siting strategies proposed in \autoref{sec:strategies} and provide details about \toolkit, our toolkit that will be publicly released, and includes the tools and datasets used in this work.

\vspace{-0.1cm}
\subsection{Experimental Setup}
\label{subsec:experiment}
{\color{blue}The problem of solar siting can be modeled as a multi-objective optimization problem, where the objectives are carbon offset, energy capacity addition, and equity metrics, and decisions variables are e.g., ZIP codes, strategies to select ZIP codes, or panel sites. Future work may make use of other multi-objective optimization techniques to produce Pareto-optimal placements, although a rigorous formulation of this optimization problem would require choices of the relative importance of each objective. These weightings could be considered as part of the optimization instance or could be chosen, but chosen weights are outside of the scope of this work.
In this work, we implement the strategies introduced in \autoref{sec:strategies} in our simulations.}  Each simulation calculates the impact of adding $N$ panels nationwide, where $N$ is chosen from a certain range and each panel's \textit{location} is dictated by one of the siting strategies as a ZIP code.  

\vspace{-0.1cm}
\subsubsection{\textbf{Projections of Future Installations}} \label{sec:projections}
According to a recent report by the Solar Energy Industries Association (SEIA)~\cite{SEIA_2024}, the number of existing solar installations in the United States is projected to double to $10$ million by 2030 and triple to $15$ million by 2034.  Furthermore, they estimate that a quadrupling to $20$ million by 2030 would be sufficient to reach net zero carbon emissions nationwide.
In our simulations, we set $N$ to a range of values based on these projections. 
Our data includes coverage of approximately 500,000 existing PV panels, so we consider $N \in$ \{0, \ 1$\times$10$^5$, \ 2$\times$10$^5$, \dots, 18$\times$10$^5$\} to cover these doubling and tripling scenarios, along with in-between scenarios.  

\subsubsection{\textbf{Impacts}}

To holistically evaluate the impact of each siting strategy, we report several quantities across all simulations.  First, we report the \textit{carbon offset}, which uses the estimates discussed throughout the paper to quantify the amount of carbon emissions (in metric tons) prevented by each strategy's panel placements.  We also report the \textit{annual electricity generation} (in kWh), which quantifies how much additional electricity will be generated by a strategy's panel placements over the course of a typical year.  Finally, we also show the equity impacts of each siting strategy by measuring the impact on the demographic inequity of both Realized Potential and Carbon Offset Potential shown in \autoref{fig:inequity-analysis}.

\subsection{\toolkit Toolkit}

To facilitate future work at the intersection of rooftop PV, decarbonization, and demographics analysis, we release the tools and data used in our analysis and experiments as an open-source toolkit at \url{https://github.com/coopersigrist/SunSight}.\\
%We hope this toolkit finds usage by those to whom this data and accompanying visualizations are inaccessible. 
\toolkit is used for all of the analysis in this paper and is designed to accommodate more general analysis and visualization. While the \toolkit toolkit currently has basic features, we plan (and encourage the research community) to actively expand its scope and add new and updated data sources.  
In what follows, we describe the most important features of \toolkit and basic usage information.

\subsubsection{\textbf{Data Scraping}}
The \toolkit toolkit is able to scrape each of the datasets mentioned in \autoref{sec:analysis}. Each dataset is scraped at the finest granularity available -- this includes a house-level granularity for Project Sunroof, although a specific home address (i.e., location) must be provided. An example usage of this is provided in \texttt{Data\_scraping/util.py}, along with a number of other data scraping scripts and examples. % to convert into longitude and latitude. 
\toolkit can also scrape data that requires access to the Project Sunroof (Google Solar API) and the Census API when provided with API keys for both (not included).  Scraping Census data for any desired demographic is supported by using the U.S. Census Bureau's unique code lists~\cite{censusCodeLists}.
%though through the census codes used by the dataset. 
We provide and scrape a number of these codes (i.e., for the demographics and variables considered in the paper) by default.
%but it is impractical to scrape them all by default.

\subsubsection{\textbf{Data Cleaning and Compilation}}
The data cleaning and compilation procedure described in \autoref{sec:analysis} is performed by a script included in the toolkit. This can be used if a different set of demographics or granularity is required. 
The cleaned and compiled datasets used in this work are provided as CSV files that may be used directly. 
Cleaned, state-level granular data is provided under \verb|Visualization/Clean_Data/data_by_state.csv|. This data includes all available data from Project Sunroof at a state-level granularity as well as each of the default census tract demographics averaged across their states. Political affiliation and energy generation data, described in \autoref{sec:analysis}, are included as well. The cleaned, ZIP code-granular data is likewise provided in \verb|Visualization/| \verb|Clean_Data/data_by_zip.csv|. This includes both the Project Sunroof and census tract data, but does not include the political affiliation or energy generation data. Each of the cleaned datasets includes data only from ZIP codes in which both the census tract and Project Sunroof data are available at ZIP code granularity, and all unusable data is removed. Versions of each of these datasets, at all available granularities, in their unmodified form can be found in the Data\/ folder, along with zips.csv, which contains all ZIP codes that are used in this work. 

\subsubsection{\textbf{Visualizations}}
{\color{blue} \toolkit's visualization component provides various ways to plot trends and observe relationships in the existing or newly scraped datasets.  
}
In particular, the toolkit supports the following types of plots ``out of the box'', with minimal additional coding required:
% The visualization portion of the toolkit is the most robust and adaptable. Included in the toolkit are the following types of visualizations:
\begin{itemize}[leftmargin=*]
    \item Complex scatter plots that can partition input ZIP codes by any feature of the data (e.g., separating high-income and low-income ZIP codes) with each partition identified by color and fit by an arbitrary polynomial or exponential function of the user's specification. For an example of a plot created using this function, see \autoref{fig:co_vs_existing_intro}. These scatter plots also have default settings to create feature-comparing plots such as \autoref{fig:carbon-energy-potential}
    \item Bar plots are similarly able to be partitioned and have a number of default settings to compare features of the data along demographic splits at any granularity (e.g., see \autoref{fig:inequity-analysis} for an example).
    \item Maps of the continental United States are supported for any data recorded at a ZIP code-level or state-level granularity. See \autoref{fig:co_per_cap_map} and \autoref{fig:state maps} for examples of each of these, respectively.
\end{itemize}

% \rev{Many other visualization modes are included in the toolkit as well.} \adaml{this sentence should go after the list.}

\subsubsection{\textbf{Evaluation Framework}}
% \rev{
The main motivation of \toolkit is to serve as an evaluation framework to estimate the long-term impact of different nationwide solar deployment strategies. 
For example, each sitting strategy presented in \autoref{sec:projections} can be executed as a single function call. 
While the results in the next section serve as the most notable representative observations, a wide variety of such results and plots are provided in \texttt{plot\_creation.py} as additional examples. 
We also provide a framework for creating and simulating new policies and their effect on key statistics such as energy efficiency and carbon offset potential.
Along with the policies simulated in \autoref{fig:Co_projection} and \autoref{fig:Cap_projection}, we provide more generic policies, such as a class of weighted multi-objective policy that can be customized for different priorities across multiple objectives.
All such policies can be found in \texttt{visualization/projections\_util.py}. Lastly, the usage of these policies and all visualizations related to policy simulation and projection from this paper can be found in \texttt{visualization/projections.py.}
% }   

%     \vspace{-0.3cm}
%     \caption{Simulated projections of adding up to 1.8 million panels according to the strategies described in \autoref{sec:strategies}. 
%     }
%     \label{fig:inequity-analysis}
%     \vspace{-0.4cm}
% \end{figure*}

\section{Experimental Evaluation}
\label{sec:evaluation}
In this section, we evaluate our siting strategies described in \autoref{sec:strategies} that seek to address the carbon inefficiencies and inequities observed in \autoref{sec:analysis} in the current siting strategy.  
We evaluate all the siting strategies using three metrics: (i) energy impact, (ii) carbon impact, and (iii) equity impact.

\subsection{Energy Impact}
% \adaml{pasted from outline: We could do a comparison between the greedy energy and equitable decarb strategies in a number of panels added vs Energy Gen plot but this would show the equitable strategy generates less power.}

\begin{figure}[t]
    % \vspace{-0.5em}
    \centering
    \includegraphics[width=\columnwidth]{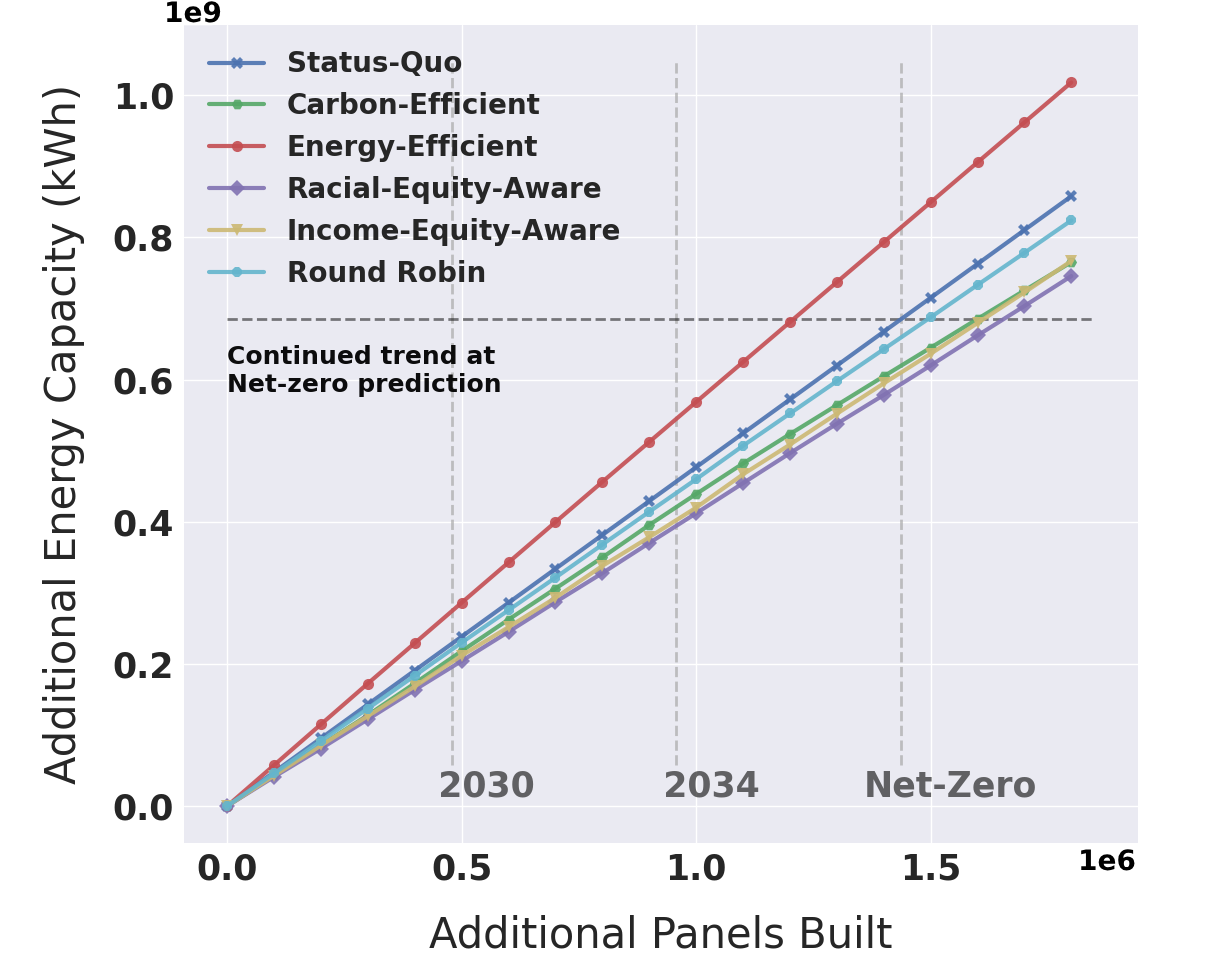}
    \vspace{-0.6cm}
    \caption{Projections of the additional (yearly) energy generation achieved using different panel siting strategies.}
    \label{fig:Cap_projection}
    \vspace{-0.5em}
\end{figure}

In \autoref{fig:Cap_projection}, we plot projections of the additional energy generation (in kWh) for each siting strategy when adding up to $1.8$ million panels to the areas covered in the dataset of \toolkit.  
Vertical grid lines in the background correspond to 2030, 2034, and net-zero scenarios discussed in \autoref{sec:projections}, and the horizontal grid line shows the additional energy generation attained by continuing the status quo till net-zero, i.e., until the number of residential solar panels is quadrupled.

The energy-efficient strategy maximizes sunlight exposure and yields the highest energy generation, surpassing all other methods as more panels are added. This is expected since the strategy targets regions with the best solar conditions. The Status Quo strategy, which maintains the current distribution patterns, performs second-best, reflecting the fact that many existing installations are already in high-sunlight regions, as we observe in \autoref{fig:state maps}(f). The ``Round Robin'' strategy, designed to ensure more equitable distribution across all factors, follows closely, achieving 94.6\% of the additional energy generation of the Status Quo, demonstrating that this strategy is both equitable and relatively efficient.

On the other hand, the siting strategies that focus only on equity objectives, such as racial-equity-aware and income-equity-aware, show slightly lower energy generation performance but remain competitive. These approaches prioritize a more equitable distribution of solar installations across different demographic groups, which may lead to lower overall energy generation due to installations in areas with less sunlight. While targeting carbon reduction, the carbon-efficient strategy also achieves respectable energy efficiency, as it likely balances between regions with high solar potential and those with high grid carbon intensity. Overall, while the energy-efficient strategy maximizes output, equity-aware strategies show that there is only a modest trade-off between energy generation and equitable distribution.

% \mo{add takeaway block to this similar to 6.2 and 6.3.}
\vspace{0.1cm}
\noindent
\emph{\textbf{Key Takeaway.} Due to abundant sunlight in states that have the bulk of existing installations, continuing the ``Status Quo'' distribution of installations results in high added generation capacity -- other siting strategies perform worse.  Our multi-objective ``Round Robin'' approach that equally weights all four factors is a close second to the ``Status Quo'' in terms of energy.}

% yields good outcomes in terms of the added annual energy generation, but suboptimal results in terms of short-term climate impact (i.e., carbon offsets).  Simple strategies such as biasing installations towards areas with proportionally high Black or low-income populations significantly outperform the status quo in this regard.  Our combination ``Round Robin'' approach that equally weights all four factors represents the ``best of both worlds'', achieving comparable additions in energy generation while maintaining a sizable advantage in carbon offsetting.}

% Intuitively, the siting strategy that optimizes for energy efficiency (i.e., average sun) exhibits the best performance in terms of additional generation.  The next best strategy is the continuation of the status quo -- this also makes intuitive sense since, as we observe in \autoref{fig:state maps}(f), the existing distribution of installations is skewed towards states in the western US that enjoy high average sunlight.  Our multi-objective ``Round Robin'' strategy is a close third in these results, achieving 94.6\% of the added yearly energy generation of the status quo continuation.

\begin{figure}[t]
% \vspace{-0.5em}
    \centering
    \includegraphics[width=\columnwidth]{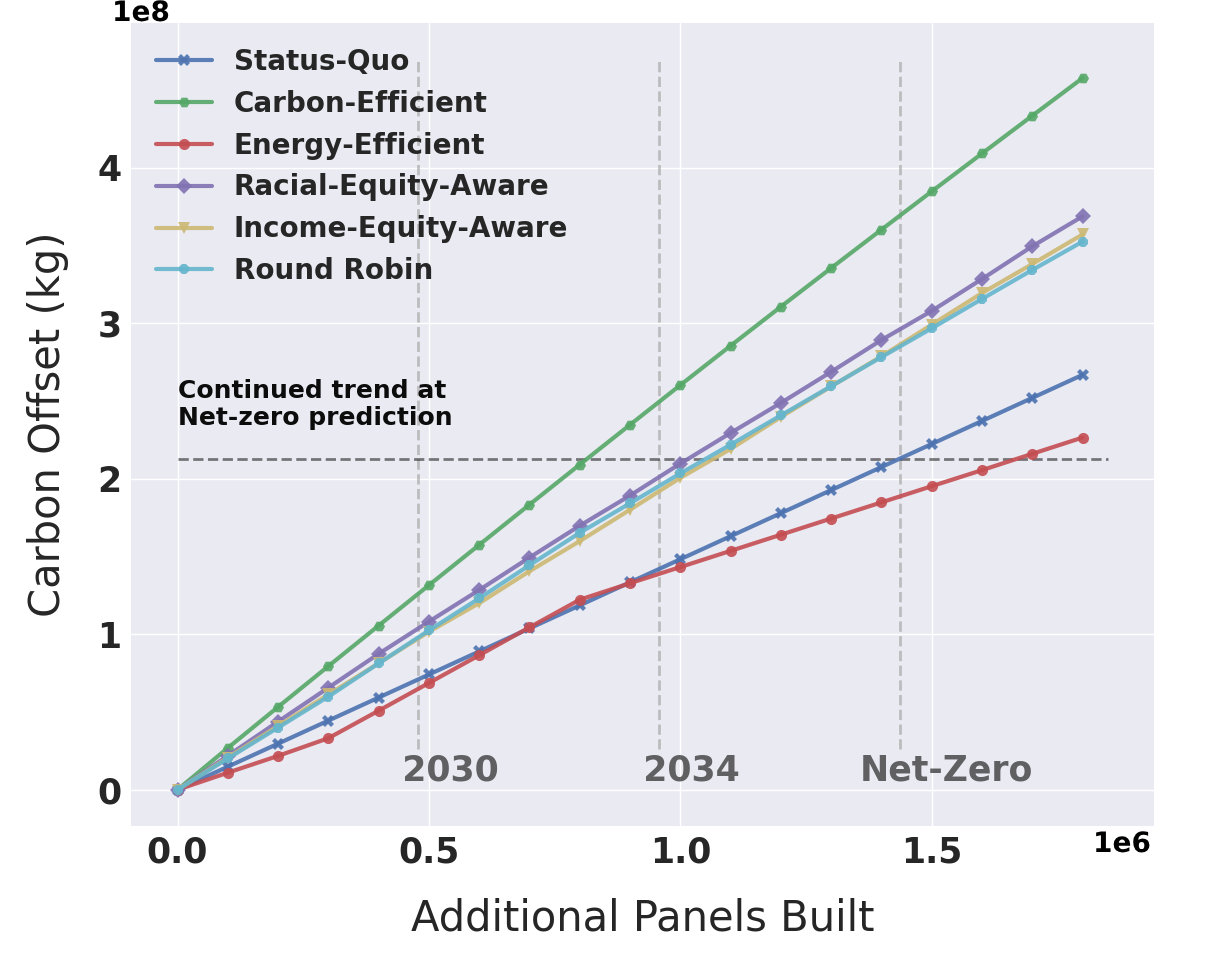}
    \vspace{-0.6cm}
    \caption{Projections of the estimated carbon emissions offset achieved using different panel siting strategies.}
    \label{fig:Co_projection}
\vspace{-0.5cm}
\end{figure}

\subsection{Carbon Impact}

In \autoref{fig:Co_projection}, we plot projections of the additional energy generation (in kWh) for each siting strategy when adding up to $1.8$ million panels to the areas covered in our dataset.  
Vertical grid lines again correspond to the scenarios discussed in \autoref{sec:projections}, and the horizontal grid line corresponds to the CO$_2$ emissions (in kg) that are offset by continuing the Status Quo of rooftop PV distribution until the net-zero scenario is attained (i.e., until the number of residential PV panels is quadrupled).

Unsurprisingly, the siting strategy that optimizes for carbon efficiency (i.e., carbon offset) exhibits the best performance in terms of additional carbon offsetting.  Surprisingly, the continuation of the Status Quo PV distribution does quite poorly in terms of carbon offsetting -- all strategies except for the strategy that optimizes for energy efficiency (i.e., average sun) do better in this regard.
The remaining siting strategies, which focus solely on currently underutilized demographics and areas, significantly outperform the continuation of the status quo.  This includes the multi-objective ``Round Robin'' strategy, which achieves a 39.8\% improvement over the carbon offset achieved by the continuation of the Status Quo distribution.

Interestingly, according to the SEIA projections for the net-zero scenario, our ``Round Robin'' strategy achieves the same ``net-zero'' carbon offset as the continuation of the Status Quo when only adding 69.0\% as many PV panels -- since doubling and tripling the amount of rooftop PV in the United States would take time, this suggests that optimizing for carbon efficiency is a powerful tool towards net-zero goals.

% Lastly, we once again use SEIA projections to estimate how many rooftop PVs would be built in a net-zero emissions scenario (particularly this is net-zero in 2030). And estimate how many years sooner we could reach this goal using round-robin over following the current policy. If such a simple policy can improve our timeline this much, we invite other work so develop more advanced policies to balance these objectives. 

\vspace{0.1cm}
\noindent
\emph{\textbf{Key Takeaway.} Adding new installations according to the existing ``Status Quo'' distribution of installations yields good energy outcomes, but suboptimal results in terms of climate impact (i.e., carbon offsets).  Simple strategies such as biasing installations towards areas with proportionally high Black or low-income populations significantly outperform the Status Quo in this regard.  The multi-objective ``Round Robin'' approach represents the ``best of many worlds'', achieving comparable additions in energy generation while maintaining a sizable advantage in carbon offsetting.}

% significantly better carbon offsets while maint

% \adaml{key takeaways for fig 6 + 7 will be formatted nicely here...}]

% Overall story of 6 + 7 : Continued is very back for carbon offset and even simple greedy policy beats it, but it does better in overall energy capacity addition (which matters on a longer time scale). Our Round Robin approach has comparable energy capacity gains while maintaining an advantage in Carbon Offset.  

\begin{figure*}[t]
    \centering
    \vspace{0.3cm}
  \subfloat[Racial-Equity-Aware]{
	\begin{minipage}[c][0.2\width]{
	   0.19\textwidth}
	   \centering
	   \includegraphics[width=1\textwidth]{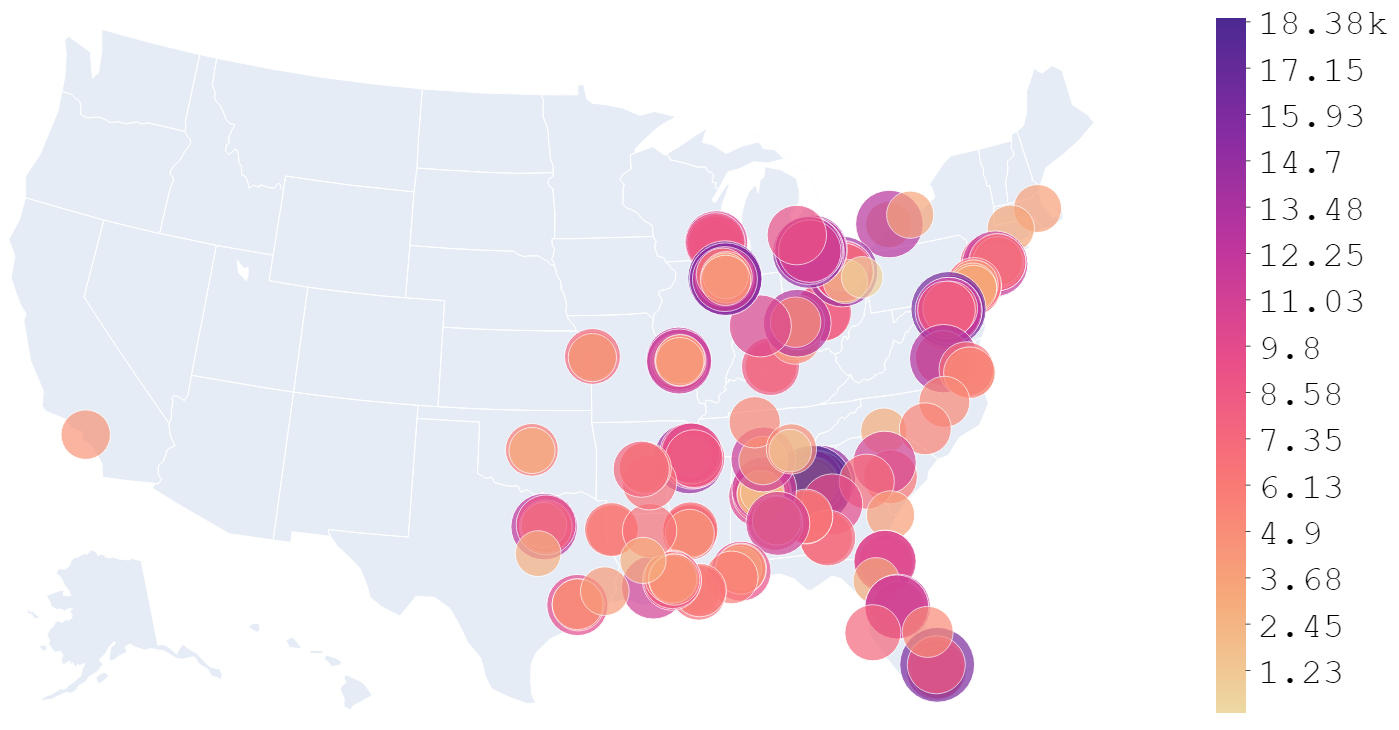}
       \vspace{0.5cm}
	\end{minipage}}
   \subfloat[Carbon-Efficient]{
	\begin{minipage}[c][0.2\width]{
	   0.19\textwidth}
	   \centering
	   \includegraphics[width=1\textwidth]{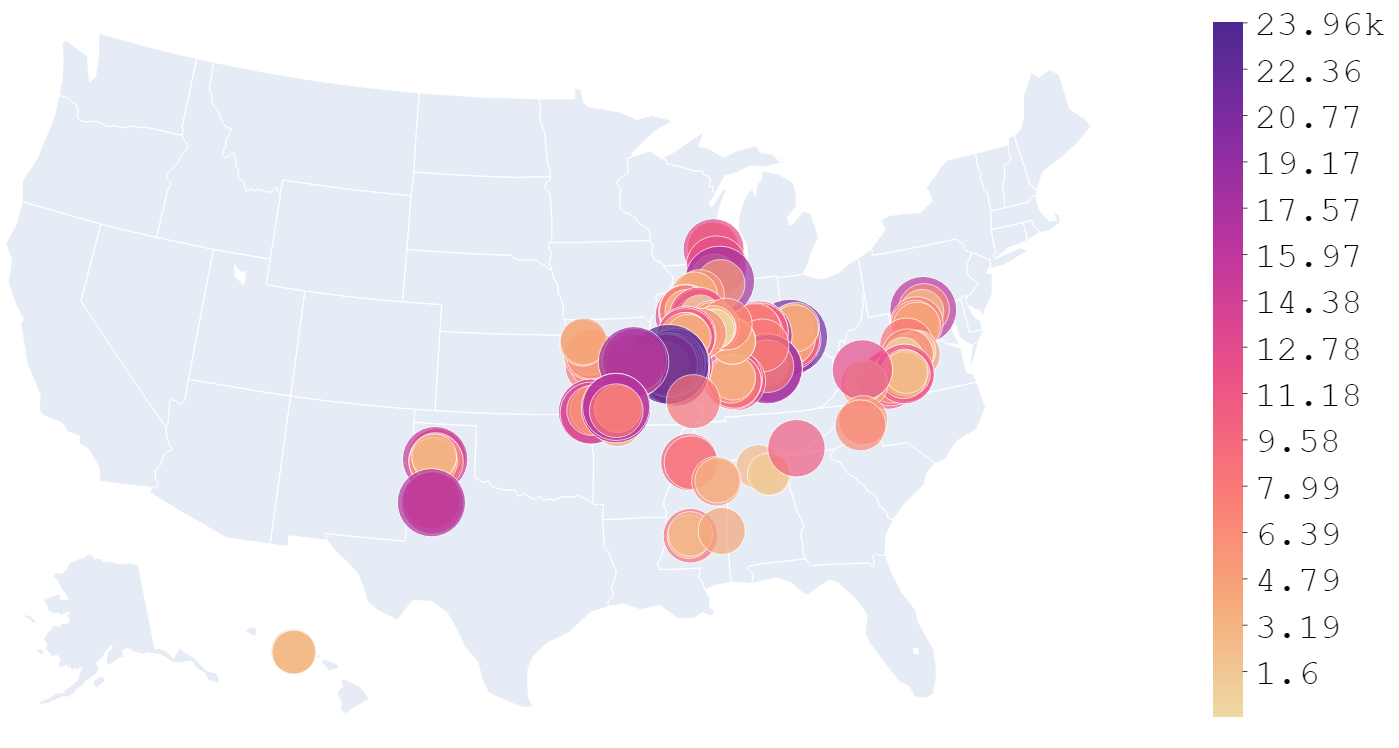}
        \vspace{0.5cm}
	\end{minipage}}
   \subfloat[Income-Equity-Aware]{
	\begin{minipage}[c][0.2\width]{
	   0.19\textwidth}
	   \centering
	   \includegraphics[width=1\textwidth]{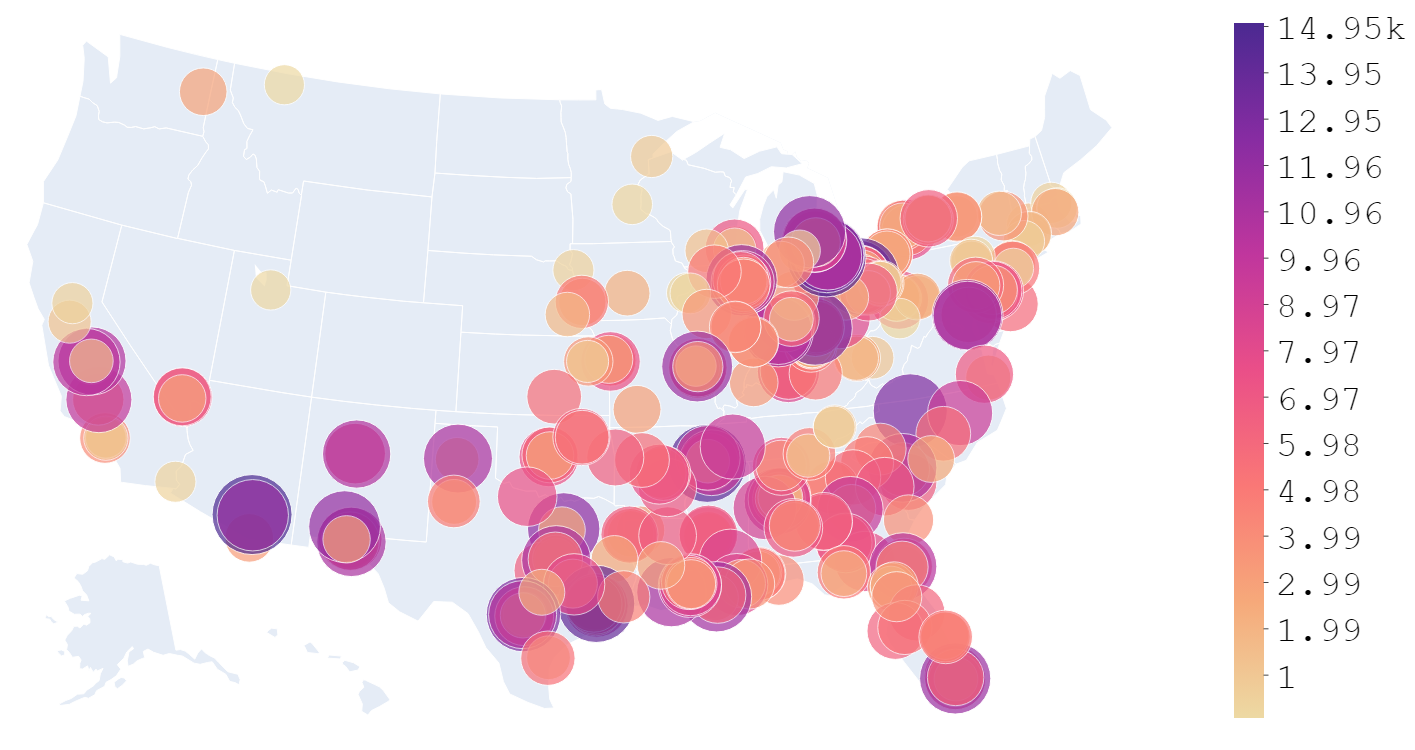}
       \vspace{0.5cm}
	\end{minipage}}
    % \hfill
      \subfloat[Energy-Efficient]{
	\begin{minipage}[c][0.2\width]{
	   0.19\textwidth}
	   \centering
	   \includegraphics[width=1\textwidth]{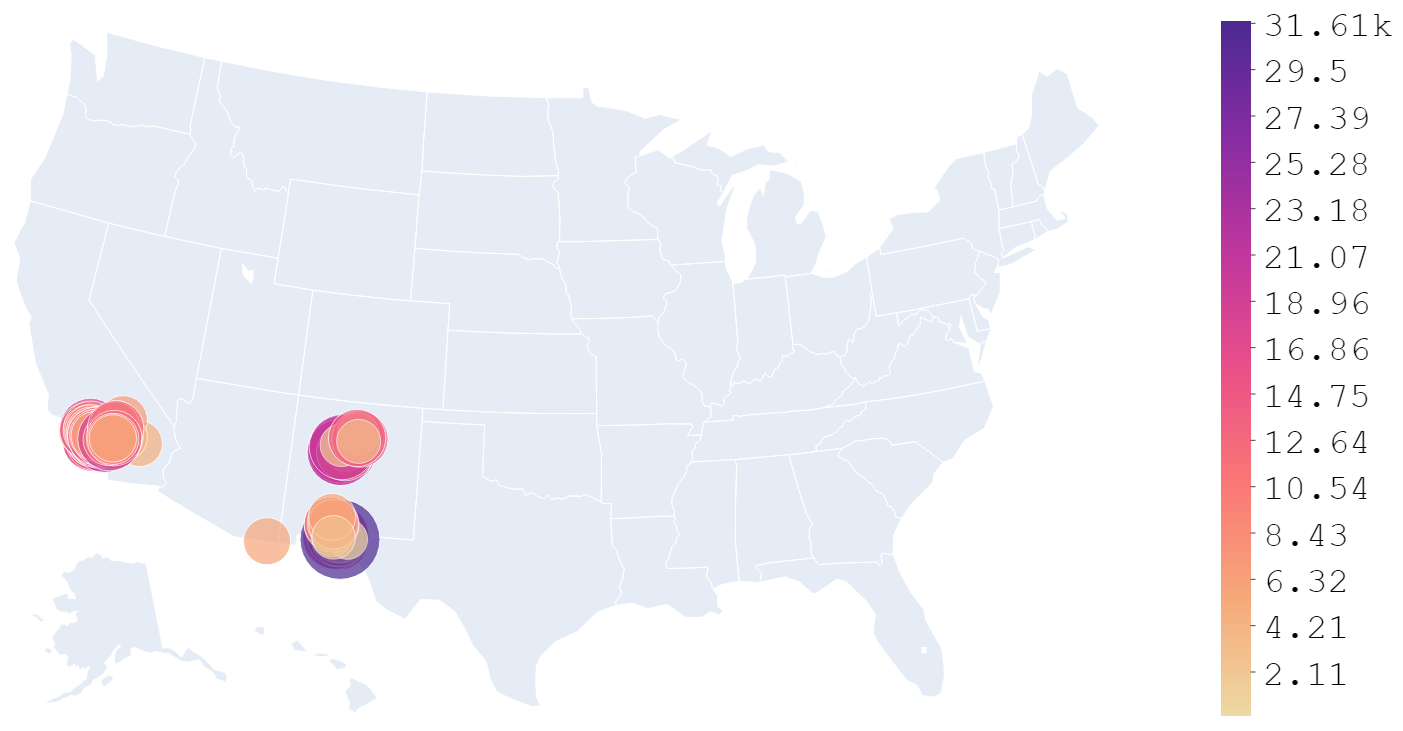}
       \vspace{0.5cm}
	\end{minipage}}
       \subfloat[Round Robin]{
	\begin{minipage}[c][0.2\width]{
	   0.19\textwidth}
	   \centering
	   \includegraphics[width=1\textwidth]{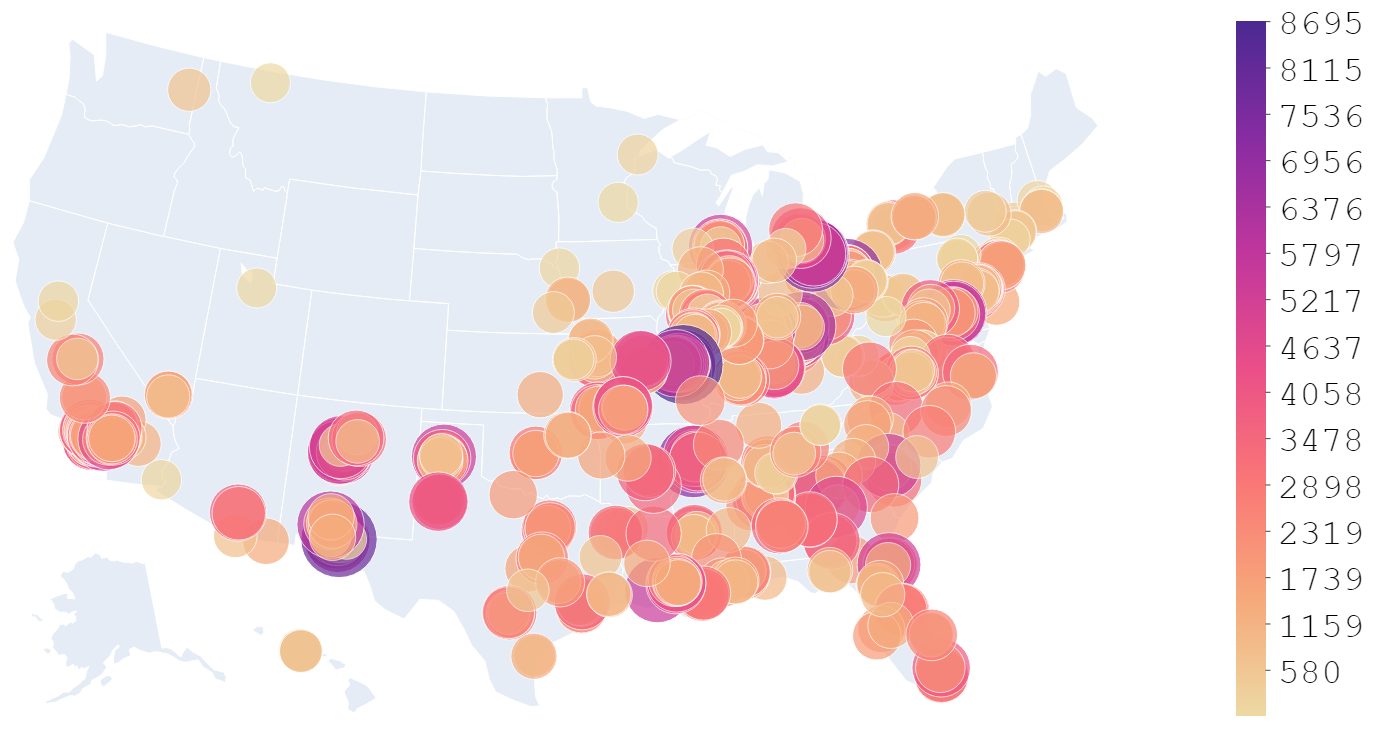}
       \vspace{0.5cm}
	\end{minipage}}
    \vspace{-1em}
    \caption{The projected locations and quantities of panels projected to be built by the strategies described in \autoref{sec:strategies}} (a) Maximizes Racial Equity of installations (namely for black population) (b) Maximizes Carbon Offset (c) Maximizes the number of panels placed in low-income ZIP codes (d) Maximizes the Energy Capacity gain and (e) is our Round Robin of each of the other strategies. Larger, darker dots represent ZIP codes which have a high number of panels placed in them (note: Round Robin has the lowest maximum).  
    \vspace{0.0cm}
    \label{fig:placement maps}
\end{figure*}

\subsection{Equity Impact}

Since many of our siting strategies optimize for socioeconomic equity factors, it is also natural to investigate their impacts on these factors \textit{after} simulating their respective siting decisions.

In \autoref{fig:placement maps}, we plot maps of the panel placement decisions made by our illustrative strategies described in \autoref{sec:strategies}. These allow us to quantify the geographic equity of each strategy.

% \begin{figure}[h]
% \centering
%     \includegraphics[width=\columnwidth]{results/Maps/Greedy_avg_sun.png}
%     \caption{Projected placement of 1.8 million panels as dictated by the siting strategy that purely optimized for energy efficiency (i.e., average sunlight).}
%     \label{fig:Greedy_energy_placement}
%     \centering
%     \includegraphics[width=\columnwidth]{results/Maps/Round Robin greedy placement (agsunset).png}
%     \caption{Projected placement of 1.8 million panels as dictated by the equity-aware ``Round Robin'' siting strategy.}
%     \label{fig:round_robin_placement}
% \end{figure}

At a high level, we observe that any siting strategy that optimizes for a single attribute will intuitively focus on small, localized regions, choosing to exhaust the possible utilization in those regions before moving to other locations. 
{\color{blue}
This is evident in \autoref{fig:placement maps}(d), where all of the simulated panels are installed in the sunny southwest while \autoref{fig:placement maps}(b) focuses on the Midwest and Coal Belt areas of the U.S. 
}

With regards to the energy-efficient \autoref{fig:placement maps}(d), as discussed previously, optimizing for average sun results in a substantially suboptimal allocation in terms of carbon offset because electric grids in places such as the southwest US are already clean during the day.

In contrast, our multi-objective ``Round Robin'' strategy, by taking ZIP codes based on several attributes, produces an allocation that is more evenly distributed and thus more realistic in practice.  In comparing \autoref{fig:placement maps}(e) with the strategy that optimizes for energy efficiency, we note that the multi-objective strategy concentrates installations in the southeast US and Midwest, areas that correspond to higher carbon offsets, above-median Black population proportions, and below-median incomes (see \autoref{fig:state maps}(e), (b), and (c), respectively). 

\begin{figure}[h]
\centering
\vspace{-0.2cm}
    \includegraphics[width=\columnwidth]{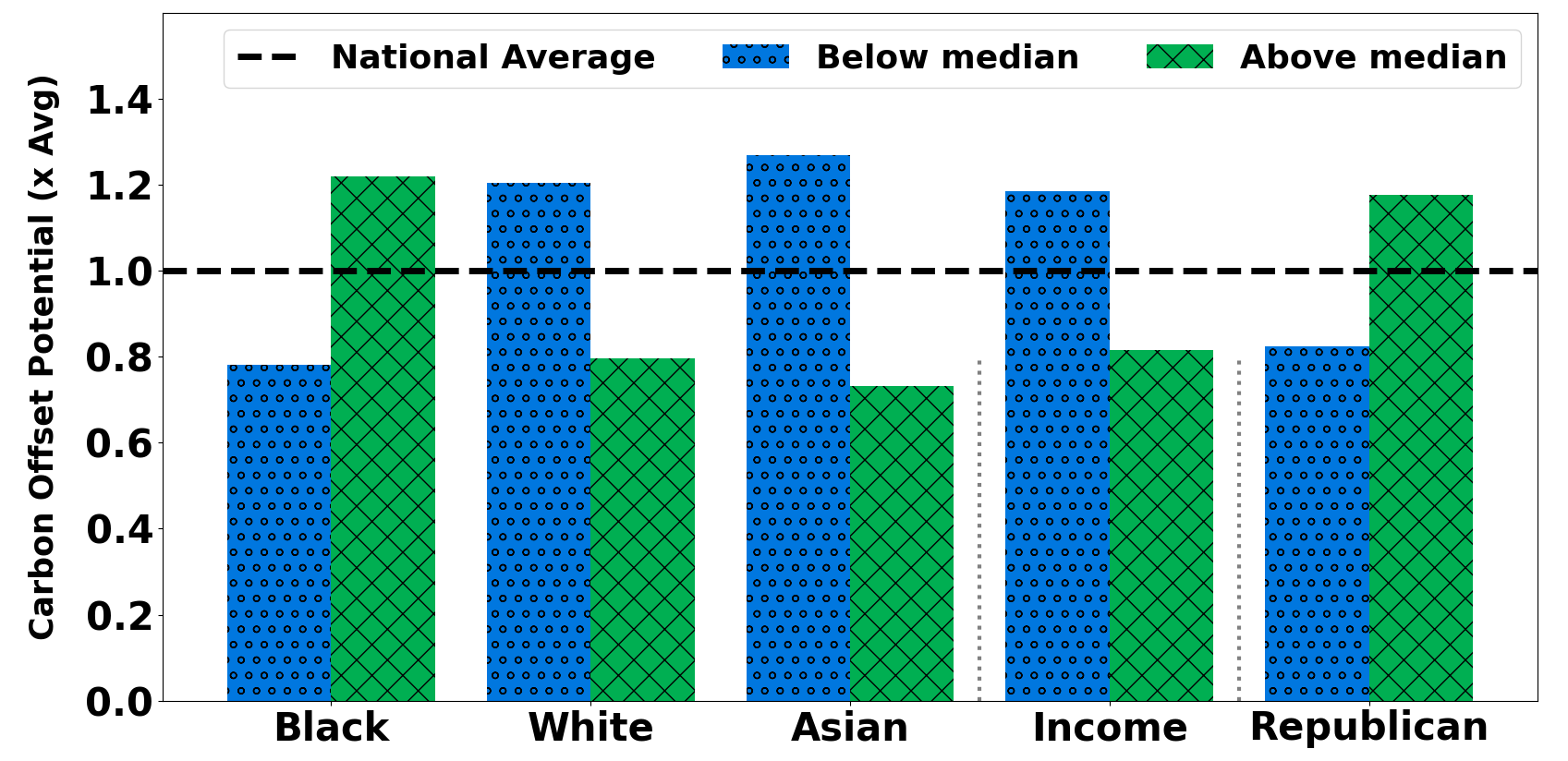}
    \centering
    \includegraphics[width=\columnwidth]{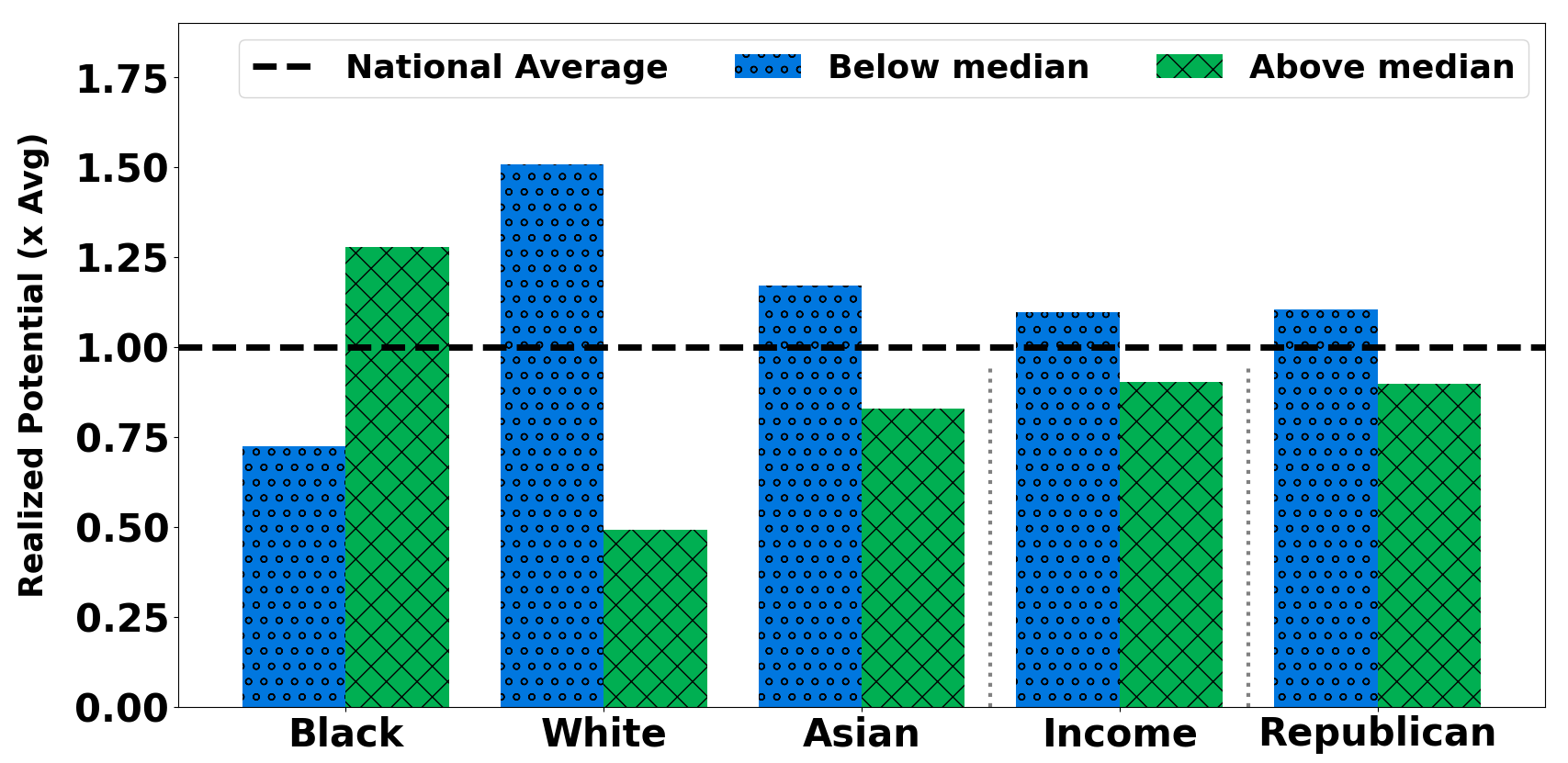}
    \caption{The projected demographic breakdown of Estimated Carbon Offset Potential and Realized Potential (as seen in \autoref{fig:inequity-analysis}) done \textit{after} the placement of 1.8 million panels via our ``Round Robin'' Strategy.} 
    \label{fig:after placement demo realized pot}
\end{figure}

In addition to its geographical equity, the ``Round Robin'' strategy addresses racial- and income-inequity too. In \autoref{fig:after placement demo realized pot} we see a much more equitable distribution of panels (i.e. Realized Potential). In Particular we see that high black population states went from an average of $\sim$50\% below the national average realized potential to having $\sim$25\% \textit{above} the national average.  

\emph{\textbf{Key Takeaway.} Compared to the Status Quo, even relatively simple policies (such as Round Robin over multiple objectives) are not only able to achieve a similar (94.6\%) energy capacity gain, but will improve carbon offsetting tremendously (39.8\%), remain geographically equitable, and substantially reduce racial- and economic-inequity.}

% \co{OUTLINE of Figure 8+9:
% \begin{itemize}
%     \item Each of these show the projected distribution of panels placed by their respective policies. These plot give us the following story: 1. Greedy policies will focus on very few regions/zips and will load them up. particularly, focusing on energy capacity potential (which is an intuitive notion, even SEIA focuses on energy capacity in their reports) will concentrate the building in areas we have shown previously are bad carbon offsetters. 2. Our round-robin policy will distribute the built panels across the country, which, for rooftop solar, is much more feasible (and cheaper to incentivize -- I don't know if we can claim that though).    
% \end{itemize}

\section{Related Work}
\label{sec:related}
\noindent In this section, we review prior work that studies adoption, incentives, and the energy generation potential of rooftop solar installations.

Residential rooftop PV adoption is relatively well-studied, particularly from the perspective of incentive programs~\cite{Sarzynski2012, Bauner2015, Hagerman2016, Crago2017, Matisoff2017, Boccard2021, Peasco2021, Kearns2022}.  
{\color{blue}
Several studies have concluded that point-of-sale rebates are more effective compared to tax credits~\cite{Crago2017, Kearns2022} -- for instance, \citet{Matisoff2017} reviewed state and utility incentives for residential PV in the United States and found that point of sale rebates were up to $8\times$ more effective compared to tax credits worth the same amount.  
}
A few studies have also considered socioeconomic equity from the frame of adoption -- for instance, \cite{Sunter:19, Dokshin2023RevisedEO} present studies using Google's Project Sunroof data to evaluate PV adoption at a census tract level, finding varying amounts of disparities along race and income lines.  
{\color{blue}
Closer to the experiments that we conduct in our case study, a few works have used simulations to estimate the impact of existing or proposed incentives on specifically adoption.  Many find that upfront subsidies encourage adoption, including~\cite{Hsu2012, Burr2016, Zander2019, Sher2022, DAdamo2022}.
}

\citet{OShaughnessy:20} consider different business models for rooftop PV and how they affect different income levels. From a policy angle, authors in~\cite{Zhou:23} compare the effectiveness of different policies designed to increase equity in rooftop PV.  %Further, \cite{Chan:17} considers the community solar model as a method to equitably distribute the benefits of PV generation.  
In another category of results~\cite{Lan:21, Ray:21, Tidwell:21} analyze the equity of PV adoption in Australia and two U.S. states (Washington, Georgia), respectively.
Finally, \citet{Crago:23} consider socio-economic equity in rooftop PV deployment and return on investment, which depends on the business model provided by the installer (i.e. whether a system is leased or owned).
Beyond adoption, \citet{REAMES2020101612} considers distributional disparities in rooftop PV \textit{energy generation} potential, analyzing at a census tract level using the REPLICA dataset~\cite{REPLICA:18}.  

While they do not consider rooftop PV, other studies have simulated the economic and climate impact of heat pumps, finding that adoption and decarbonization potential both depend heavily on e.g., electricity generation mix and incentives, aligning with our results~\cite{Barnes2020, Gaur2021, Kokoni2021, PobleteCazenave2023}.  
{\color{blue}
Similarly, several studies have examined the impacts of rooftop PV on decarbonization by quantifying the overall effects on electricity supply and generation mix -- these generally adopt a broad (i.e., national or global) perspective on the issue to understand the role of small-scale residential systems in a broader energy transition~\cite{Gernaat:20, Chapman:16, Agnew2015}.
}
% While many of our high-level findings are shared with~\cite{Reames:20}, we leverage Google's Project Sunroof data set, which provides data at the individual household level and incorporates acute shading analysis.  We also analyze demographic data at a census block group level, which is more granular than the census tract level.  %Additionally, where data is available, we consider how rooftop solar fits into the broader residential electrification equation.% by comparing potential generation against current and future energy usage.

{\color{blue}
In the context of placing solar panels, several works have naturally considered optimizing the placement of solar panels within a small area (e.g., on a building or in a campus), generally with the objective of maximizing energy generated~\cite{Zhong2020, Manowska2023, Kucuksari2014, Kuterbekov2024, Reffat2024}.  Closest to our work, \citet{ElKontar2020} consider the placement of community solar (i.e., placing PV panels on rooftops throughout a single residential community) with the dual objectives of maximizing energy generation and minimizing operational costs.
Compared to the works discussed here, ours is the first, to our knowledge, that specifically examines multi-dimensional connections between climate impact (i.e., in terms of carbon reduction) of rooftop solar PV, and distributional effects (e.g., disparate adoption correlated with socioeconomic factors).
}

\section{Conclusion}
\label{sec:related}

While prior works have established that inequities exist in the distribution of rooftop solar installations in the United States, most studies up to this point have focused on adoption (i.e., in terms of number of installations) or energy generation (in terms of returns on investment or sunlight potential).  In this work, we adopt a carbon-focused lens and consider how the socioeconomically-correlated distribution of rooftop PV throughout the US impacts the climate benefits of the technology in terms of direct CO$_2$ reduction.  We find that regions with above-median Black population proportions, below-median incomes, and Republican voting records tend to not not only have significantly fewer installations, but that future installations in those regions would offset more carbon.

To address these concerns, we propose a set of equity- and carbon-aware solar siting strategies that prioritize solar development in specific areas based on their attributes, emulating the effects of e.g., targeted incentives. In evaluating these strategies, we develop the \toolkit simulation/visualization toolkit and data set, which we are releasing publicly alongside this paper to facilitate continued analysis and experimentation in this area. Our projections indicate that a multi-objective siting strategy can enhance both distributional equity and the carbon-efficiency of current installation trends by up to 39.8\%.

\bibliographystyle{ACM-Reference-Format}
\bibliography{paper}

% \appendix

% \section{Revision Summary}
% \label{sec:revisions}
% \input{9-revision-summary}

% \section{Detailed Reviewer Responses}
% \label{sec:responses}
% \input{10-revision-detailed}

% \clearpage
% \section*{Outline}
% \input{00-outline}

\end{document}